\documentclass[a4paper,twocolumn,11pt]{quantumarticle}
\pdfoutput=1
\usepackage[utf8]{inputenc}
\usepackage[english]{babel}
\usepackage[T1]{fontenc}
\usepackage{amsmath}
\usepackage{hyperref}
\usepackage{multirow}
\usepackage{tikz}
\usepackage{lipsum}
\usepackage{color}
\usepackage{ulem}
\usepackage{graphicx,subfigure}

\begin{document}

\title{High-fidelity quantum teleportation through noisy channels via weak measurement and environment-assisted measurement}

\author{Sajede Harraz}
\affiliation{Department of Automation, University of Science and Technology of China, Hefei 230027}
\orcid{0000-0002-9708-2107}
\thanks{sajede@ustc.edu.cn}
\author{Jiao-Yang Zhang}
\affiliation{Department of Automation, University of Science and Technology of China, Hefei 230027}

\author{Shuang Cong}
\affiliation{Department of Automation, University of Science and Technology of China, Hefei 230027}
\orcid{0000-0001-8101-0128}
\maketitle

\begin{abstract}
  A perfect teleportation protocol requires pure maximally shared entangled states. While in reality the shared entanglement is drastically degraded due to the inevitable interaction with the noisy environment. Here, we propose a teleportation protocol to teleport an unknown qubit through amplitude damping channels with \textcolor{black}{a} fidelity up to one \textcolor{black}{with a single copy of the entangled state}. \textcolor{black}{Our proposed teleportation protocol, while illustrated using the Bell and W entangled states as examples, can be utilized with any type of entangled states.} In our protocol, we utilize environment-assisted measurement during the entanglement distribution\textcolor{black}{,} and further modify the \textcolor{black}{original} teleportation protocol to apply weak measurement in the last step of teleportation. \textcolor{black}{We find a balance between teleportation fidelity and success probability by varying the strength of the weak measurement.} Furthermore, we \textcolor{black}{investigate the protection of} controlled teleportation protocol\textcolor{black}{s}, where all the qubits of the \textcolor{black}{entangled} state pass through the amplitude damping channel. \textcolor{black}{In particular, for the controlled teleportation with the W state, the decoherence of the shared entanglement can be totally suppressed by using EAM, hence no weak measurement is required to achieve an average teleportation fidelity of unity. The numerical simulation results reveal that our proposed teleportation protocol outperforms both the weak measurement based probabilistic teleportation protocol and the original teleportation protocols without protection.} 
  \end{abstract}

Quantum teleportation is a quantum communication task which sends an unknown qubit from a sender to a receiver by using shared entanglement and classical communications \cite{lab1, lab2}. The original protocol proposed by Bennett $et~al.$ \cite{lab3} uses \textcolor{black}{a Bell state} as the shared \textcolor{black}{entanglement}. Other teleportation protocols are also developed by using \textcolor{black}{other multi-}qubit entangled state\textcolor{black}{s such as} GHZ states and W states \cite{lab4, lab5, lab6, lab7}. The W states are \textcolor{black}{typically preferred due to their resilience against the} loss of particles\textcolor{black}{, i.e.,} if \textcolor{black}{one of the particles in the W state is traced out}, then there \textcolor{black}{remains significant} genuine entanglement between the remaining two. \textcolor{black}{Typically, the sender or receiver prepares and sends the shared entangled state to one another}, which has been extensively studied \cite{lab8, lab9, lab10, lab11}. \textcolor{black}{Conversely, in} controlled teleportation\textcolor{black}{, a crucial aspect of secure quantum communication, a third party prepares and sends the shared entangled state to both} the sender and receiver \cite{lab12, lab13, lab14, lab15, lab16}. As the controller of the whole teleportation protocol, the third party can terminate the teleportation process in time when he notices something aberrant or insecure. 

In any realistic implementation of the teleportation protocol, noise is unavoidably present and affects the entangled state during its transmission to teleportation parties \cite{lab17, lab18}. Therefore, the entanglement degree of the quantum channel degraded, which seriously deteriorates the performance of the teleportation \cite{lab19, lab20}. \textcolor{black}{Entanglement purification is proposed to improve the fidelity of teleportation in presence of noise, which enhances the entanglement of a pair of qubits at the expense of numerous identically prepared entangled qubits in combination with local operations and the classical communication \cite{lab21, lab22, lab23, lab24}. Quantum error correction is another method for protecting qubits as they are transmitted through a noisy channel, but it also requires a large number of redundant qubits to encode logical quantum information \cite{lab25, lab26, lab27, lab28, lab29, lab30, lab31}.} Recently, \textcolor{black}{weak measurement-based decoherence control strategies have gained popularity and been verified both theoretically and experimentally; see Ref.~\cite{lab32} and the references therein. In Ref.~\cite{lab8}, the teleportation process is analyzed in the framework of quantum measurement and its reversal (MR), where the well-designed weak measurement operators are applied in the last step of the teleportation protocol to overcome the effects of the noisy channel. However, the performance of the MR framework is dependent on the intensity of the noise, and the success probability of achieving high-fidelity teleportation in the presence of intense decay rates dramatically decreases.} 

\textcolor{black}{All the above schemes are applied on the state of the system, while there are some schemes which can effectively protect the system state by directly manipulating the noisy channel, such as environment-assisted error correction (EAEC) \cite{lab33}. In the EAEC scheme, a measurement is applied to the noisy environment coupled to the system of interest, followed by restoration operations on the system conditioned on the results of the measurement on the environment. In this scheme, all the Kraus operators must be proportional to unitary operators; hence, by applying a reversal operation conditioned on the outcome of the measurement performed on the environment, the initial unknown state of the system is recovered. It is shown that the success probability and the fidelity of the recovered state in the EAEC scheme are always equal to 1. Later, a probabilistic extension of EAEC by combining environment measurement and weak measurement is proposed for the noisy channels with non-random unitary decompositions \cite{lab34}. In this scheme, just some of the Kraus operators should be invertible instead of unitary. The idea is to perform a measurement to the environment coupled to the system of interest, and select the system states corresponding to the invertible Kraus operators of the noisy environment. Afterwards, by applying the designed weak measurement reversal operators, the initial state of the system is recovered \cite{lab35, lab36, lab37}. }

In this \textcolor{black}{article}, we propose a teleportation protocol by utilizing EAM and WM (TP-EW), to transmit an unknown qubit through noisy channels \textcolor{black}{via a single copy of an entangled state}\footnote{\textcolor{black}{Matlab codes for regenerating the results of this article are available from <https://github.com/Sjd-Hz/High-fidelity-TP-EW>.}}. \textcolor{black}{We provide the detailed procedure of the modified teleportation protocols with W and Bell entangled states, but it should be stressed that the proposed teleportation protocol is applicable to any type of entanglement. In addition, }the TP-EW is applicable for teleportation through arbitrary decoherence channels with at least one invertible Kraus operator. \textcolor{black}{In this article, we only consider the amplitude damping channel (ADC) and prove that the considered Kraus operators are the optimal decomposition in the sense of maximizing the success probability.} First, we assume that Alice (sender) prepares \textcolor{black}{the entangled state}, and sends one qubit to Bob (receiver) through \textcolor{black}{an ADC}. \textcolor{black}{The receiver applies EAM on the ADC during the entanglement distribution process. The teleportation process will not begin unless the ADC is detected to be in the unexcited state. Following a successful entanglement distribution via EAM, the teleportation process begins, with the receiver using the well-designed weak measurement operators in the last step to recover the input state at his end. To strike a balance between the average teleportation fidelity and the total teleportation success probability, we define the weak measurement strength as a variable and discuss teleportation performance for various amounts of weak measurement strength.} We show that, by considering designed weak measurement operators, the proposed TP-EW is able to achieve teleportation fidelity equal to one independent of the \textcolor{black}{intensity of the noise. Subsequently, we investigate the application of TP-EW to controlled teleportation with W and Bell states}, where a third party (controller) prepares the entangled state and sends the relevant qubits to Alice and Bob through \textcolor{black}{independent ADCs. Particularly, in the case of controlled teleportation with the W state through ADCs, we show that employing EAM during the entanglement distribution process is sufficient to attain an average teleportation fidelity of unity. The comparison results with the original teleportation protocol under no protection and a pioneer weak measurement-based teleportation protocol in the MR framework demonstrate that TP-EW achieves a higher average teleportation fidelity for all decaying rates with a competitive total success probability.}

\textcolor{black}{The remainder of this article} is organized as follows. \textcolor{black}{In Section 2, we briefly review the EAM technique.} In Section \textcolor{black}{3}, we present the proposed TP-EW through noisy channels and analyze its performance in details. In Section \textcolor{black}{4}, we \textcolor{black}{investigate the application of TP-EW} in controlled teleportation protocol\textcolor{black}{s}. Finally, our conclusion is given in Section \textcolor{black}{5}.

\section{\textcolor{black}{Environment-assisted measurement}}
\textcolor{black}{In this section, we briefly explain the fundamental concept of EAM, which is a key component of our teleportation protocols.}

\textcolor{black}{According to the Schrödinger equation, the evolution of the total (system + environment) density matrix is given by ${\rho _{{\rm{tot}}}}(t) = U(t)\left( {{\rho _S}(0) \otimes {\rho _E}} \right){U^\dag(t) }$, where $U(t) = \exp \left( { - {\rm{i}}{H_{{\rm{tot}}}}t} \right)$ is the total evolution operator with the total Hamiltonian ${H_{{\rm{tot}}}}$ including the interaction between system and environment. Assuming the environment is in a vacuum state $|0{\rangle _E}$, the evolution of the reduced density matrix of the system is then obtained by tracing over the environmental degrees of freedom as \cite{lab38}}
\begin{equation}\label{Eq1}
\begin{aligned}
\textcolor{black}{{\rho _S}(t)} &= \textcolor{black}{{\rm{T}}{{\rm{r}}_E}\left( {{\rho _{{\rm{tot}}}}(t)} \right)}\\
 &= \textcolor{black}{\sum\limits_n {{}_E{{\langle {\psi _n}|U(t)|0\rangle }_E}{\rho _S}(0){}_E{{\langle 0|{U^\dag }(t)|{\psi _n}\rangle }_E}} }\\
 &\buildrel \Delta \over = \textcolor{black}{\sum\limits_n {{K_n}{\rho _S}(0)K_n^\dag }} 
\end{aligned}
\end{equation}
\textcolor{black}{where ${K_n} \buildrel \Delta \over = {}_E{\langle {\psi _n}|U(t)|0\rangle _E}$ are the so-called Kraus operators that depends on the initial state of the environment and the choice of the complete basis of the environment $\left\{ {|{\psi _n}{\rangle _E}} \right\}$. Thus, changing the basis of the environment from $\left\{ {|{\psi _n}{\rangle _E}} \right\}$ to $\left\{ {|{\psi _m}{\rangle _E}} \right\}$ as ${}_E\langle {\psi _m}| = \sum\nolimits_m {\langle {\psi _m}|{V_{n,m}}}$ will lead to another set of Kraus operators ${L_m} \buildrel \Delta \over = {}_E{\langle {\psi _m}|U(t)|0\rangle _E}$, where $V_{n,m}$ is a unitary matrix, and the relation between the two set of Kraus operators is described as}
\begin{equation}\label{Eq2}
\begin{aligned}
\textcolor{black}{{K_n}} &= \textcolor{black}{\sum\limits_m {{V_{n,m}}{}_E{{\langle {\psi _m}|U(t)|0\rangle }_E}}} \\
 &= \textcolor{black}{\sum\limits_m {{V_{n,m}}{L_m}}}
\end{aligned}
\end{equation}

\textcolor{black}{After one performs a measurement on the environment, it collapses into an eigenstate of the measured observable. Subsequently, the system will also be projected into a state corresponding to each environmental state after measurement, i.e., ${\rho _{s,n}} = {K_n}{\rho _S}(0)K_n^\dag$, if the environment collapses into the $n^{\rm{th}}$ eigenstate. The Kraus decomposition in Eq.~\eqref{Eq1} is random unitary (RU) if $K_n = c_n U_n$ for each $n$, where $U_n^{\dagger} = U_n^{-1}$ and $\sum\nolimits_n {|{c_n}{|^2}}  = 1$. Hence, according to EAEC scheme \cite{lab33}, one can recover the damped state of the system by applying an reversal operation based on the environmental measurement outcome $n$, i.e., ${\rho _R} = {R_n}{\rho _{s,n}}R_n^\dag  = {\rho _S}(0)$, where ${R_n} = \left( {1/{c_n}} \right)U_n^{ - 1}$. However, if a Kraus operator is not an RU type, this scheme fails, since reversal operations are not available. Later, by combining environment measurement and weak measurement in Ref.~\cite{lab34}, the authors restored quantum states in the presence of non-RU type noise with at least one invertible Kraus operator. Thus, after a measurement is applied on the environment, only the quantum trajectories corresponding to the invertible Kraus operator are considered, and others are discarded. In the end, the initial state of the system is recovered by a weak measurement operator that is defined as the inverse of the invertible Kraus operator as} 
\begin{equation}\label{Eq3}
    \textcolor{black}{R_n = N_n K_n^{-1}}
\end{equation}
\textcolor{black}{where $N_n$ is the normalization factor given by}
\begin{equation}\label{Eq4}
    \textcolor{black}{{N_n} = \min \{ \sqrt {{\lambda _i}} \}}
\end{equation}
\textcolor{black}{with $\lambda_i$'s being the eigenvalues of the matrix $K_n K_n^{\dagger}$. }

\section{Teleportation protocol through noisy channels by utilizing environment-assisted measurement and weak measurement}
\begin{figure}[h]
    \centering
    \includegraphics[width=0.45\textwidth]{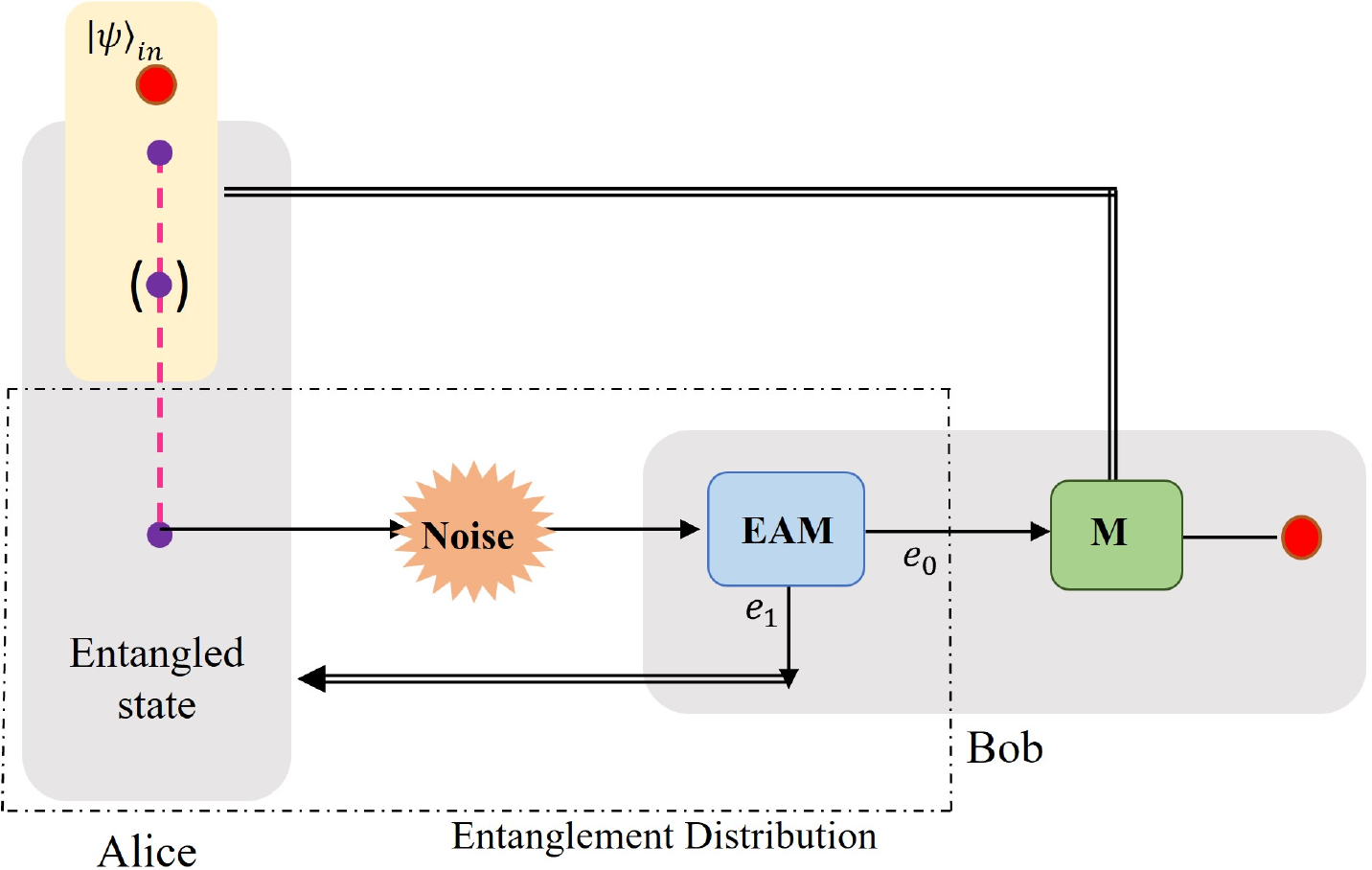}
  
\caption{The schematic diagram of proposed TP-EW. The double lines indicate the classical communications\textcolor{black}{, and the dashed line indicates quantum entanglement}.}
\label{fig1}
\end{figure}

In this section, we present the details of our proposed teleportation protocol through noisy channels by utilizing EAM and weak measurement, \textcolor{black}{where EAM is applied during the entanglement distribution process and weak measurement in the last step of the teleportation both by the receiver}. First, Alice prepares the entangled state. She keeps \textcolor{black}{her part of the entangled state} and sends one qubit to Bob through \textcolor{black}{an ADC}. \textcolor{black}{Next, Bob performs a measurement on the noisy channel. His objective} is to monitor the noisy channel, keep the system states corresponding to invertible Kraus operators of the channel and discard \textcolor{black}{others} corresponding to non-invertible Kraus operators. Hence, \textcolor{black}{after a successful EAM,} we can design weak measurement operators to be applied in the last step of the teleportation, to cancel the effects of the noise and retrieve the input state of the teleportation at the receiver's \textcolor{black}{end}. \textcolor{black}{Our proposed teleportation protocol is applicable to quantum teleportation protocols with all kinds of shared entanglement. Here we consider teleportation protocols with W and Bell entangled states through an ADC, and demonstrate that the fidelity and success probability of the proposed TP-EW are the same for both types of entangled states.} The schematic diagram of our proposed TP-EW is given in Fig. \ref{fig1} and the detailed procedure is given as follows.

\subsection{\textcolor{black}{Teleportation with W state}}
\textcolor{black}{In what follows, we elaborate the details of the proposed protocol for W-type entangled states \cite{lab39}. }

Alice has an unknown qubit and wishes to teleport to Bob as
\begin{equation}\label{Eq5}
\textcolor{black}{|{\psi _{{\rm{in}}}}\rangle  = \alpha |0\rangle  + \beta |1\rangle }
\end{equation}
where $|\alpha {|^2} + |\beta {|^2} = 1$.

Alice prepares \textcolor{black}{a shared entangle state from a class of W state as} \cite{lab15}:
\begin{equation}
\label{Eq6}
\begin{aligned}
|{{\text{W}}}{\rangle _{123}} =& \frac{1}{{\sqrt {2 + 2n} }}{(|100\rangle _{123}}\\
&+ \sqrt n {e^{i\gamma }}|010{\rangle _{123}} + \sqrt {n + 1} {e^{i\delta }}|001{\rangle _{123}})
\end{aligned}
\end{equation}
where \textcolor{black}{$n \in {\bf{R}}^+$, and $ \gamma, \delta \in {\bf{R}}$ are global} phases. \textcolor{black}{For the sake of simplicity, we set $n = 1$ and $\gamma  = \delta  = 0$ hereafter. }  

She keeps the qubits 1 and 2, and sends the qubit 3 of the entangled shared state to Bob through an ADC. The ADC is defined by the well-known Kraus operators as 
\begin{equation}
\label{Eq7}
e_0 = \left[ {\begin{array}{*{20}{c}}
1&0\\
0&{\sqrt {1 - r} }
\end{array}} \right],~e_1 = \left[ {\begin{array}{*{20}{c}}
0&{\sqrt r }\\
0&0
\end{array}} \right]
\end{equation}
where \textcolor{black}{$r \in [0,1]$} is the \textcolor{black}{magnitude of the decoherence and represents the probability of decay from the upper level $|1\rangle $ to the lower level $|0\rangle $ with $r = 1 - \exp \left( { - \Gamma t} \right)$, in which $\Gamma$ is the energy relaxation rate and $t$ is the evolving time. The Kraus decomposition of the noisy channel is non-trivial, since it is closely related to the success probability of the proposed TP-EW. The proof of the optimality of this Kraus decomposition is given in Appendix A.}

Then Bob performs the EAM \textcolor{black}{on the ADC} and tells the result to Alice. If the channel is in \textcolor{black}{an} unexcited state ($e_0$), the entanglement distribution is successfully done, and they can start the teleportation process. Otherwise, he discards the entanglement distribution at this time and restarts the process. Since only the third qubit of the shared entangled state passes through the noisy channel, the applied Kraus operators for \textcolor{black}{the W state in Eq.~\eqref{Eq6} are constructed as}
\begin{equation}
\label{Eq8}
{E_0^{\text{W}}} = I_2 \otimes I_2 \otimes e_0,~{E_1^{\text{W}}} = I_2 \otimes I_2 \otimes e_1
\end{equation}
where $I_2 = [1,0;0,1]$ is the \textcolor{black}{$2 \times 2$} identity operator\textcolor{black}{, and $e_i$'s are the Kraus operators of the ADC in Eq.~\eqref{Eq7}}.

By considering three-qubit Kraus operators in Eq. \eqref{Eq8}, we only keep the \textcolor{black}{quantum trajectories} corresponding to ${E_0^{\text{W}}}$. \textcolor{black}{Hence, after a successful EAM, }the quantum channel between two partners has been constructed, and the \textcolor{black}{normalized state of the } shared \textcolor{black}{entanglement} is described as
\begin{equation}\label{Eq9}
\begin{aligned}
\textcolor{black}{|{{\rm{W}}^{E_0^{\rm{W}}}}{\rangle _{123}}} =& \textcolor{black}{\frac{{E_0^{\rm{W}}|{\rm{W}}{\rangle _{123}}}}{{\sqrt {{}_{123}{{\langle {\rm{W}}|{{(E_0^{\rm{W}})}^\dag }E_0^{\rm{W}}|{\rm{W}}\rangle }_{123}}} }}}\\
 =& \textcolor{black}{\frac{1}{{\sqrt {4 - 2r} }}{(|100\rangle _{123}} + |010{\rangle _{123}}}\\
 &\textcolor{black}{+ \sqrt {2(1 - r)} |001{\rangle _{123}})}
\end{aligned}
\end{equation}
\textcolor{black}{And the success probability of the entanglement distribution via EAM is}
\begin{equation}\label{Eq10}
    \textcolor{black}{g_{{\rm{EAM}}}^{\rm{W}} = {}_{123}{\langle {\rm{W}}|{(E_0^{\rm{W}})^\dag }E_0^{\rm{W}}|{\rm{W}}\rangle _{123}} = 1 - \frac{r}{2}}
\end{equation}

\textcolor{black}{Note that the teleportation process can be started as long as a successful entanglement distribution via EAM has been achieved. Hence, we use the normalized shared entangled state in Eq.~\eqref{Eq9}.} 

To start the teleportation, Alice interacts the input qubit with her qubit of the entangled shared state. \textcolor{black}{Thus, t}he state of the \textcolor{black}{total} system \textcolor{black}{consisting of the input qubit in Eq.~\eqref{Eq5} and the shared entanglement in Eq.~\eqref{Eq9}} becomes
\begin{equation}\label{Eq11}
\begin{aligned}
\textcolor{black}{|\psi _{{\rm{tot}}}^{E_0^{\rm{W}}}\rangle}  =& \textcolor{black}{|{\psi _{{\rm{in}}}}\rangle  \otimes |{{\rm{W}}^{E_0^{\rm{W}}}}{\rangle _{123}}}\\
 =& \textcolor{black}{\frac{1}{{\sqrt {4 - 2r} }}}{[\alpha {(|010\rangle _{{\rm{\textcolor{black}{in},1,2}}}} + |001\rangle _{{\rm{\textcolor{black}{in},1,2}}}})|0{\rangle _3}\\
 &~~~~~~~~~~~+ \sqrt {2(1 - r)} \alpha |000{\rangle _{{\rm{\textcolor{black}{in},1,2}}}}|1{\rangle _3}\\
 &~~~~~~~~~~~+ \beta {(|110\rangle _{{\rm{\textcolor{black}{in},1,2}}}} + |101{\rangle _{{\rm{\textcolor{black}{in},1,2}}}})|0{\rangle _3}\\
 &~~~~~~~~~~~+ \sqrt {2(1 - r)} \beta |100{\rangle _{{\rm{\textcolor{black}{in},1,2}}}}|1{\rangle _3}]\\
 \buildrel \Delta \over =& \textcolor{black}{\frac{1}{{\sqrt {4 - 2r} }}}{[|\textcolor{black}{{\eta _1}}\rangle _{{\rm{\textcolor{black}{in},1,2}}}}{(\alpha |0\rangle _3} + \beta \sqrt {1 - r} |1{\rangle _3})\\
 &~~~~~~~~~~~+ |\textcolor{black}{{\eta _2}}{\rangle _{{\rm{\textcolor{black}{in},1,2}}}}{(\alpha |0\rangle _3} - \beta \sqrt {1 - r} |1{\rangle _3})\\
 &~~~~~~~~~~~+ |\textcolor{black}{{\eta _3}}{\rangle _{{\rm{\textcolor{black}{in},1,2}}}}{(\beta |0\rangle _3} + \alpha \sqrt {1 - r} |1{\rangle _3})\\
 &~~~~~~~~~~~+ |\textcolor{black}{{\eta _4}}{\rangle _{{\rm{\textcolor{black}{in},1,2}}}}{(\beta |0\rangle _3} - \alpha \sqrt {1 - r} |1{\rangle _3})]
\end{aligned}
\end{equation}
where \textcolor{black}{the subscript ``in'' denotes the input qubit in Eq.~\eqref{Eq5}}, \textcolor{black}{$\left\{ {\left. {|{\eta _i}\rangle } \right|i = 1,2,3,4} \right\}$ is a complete set of orthogonal bases with}
\begin{equation}\label{Eq12}
\begin{aligned}
\textcolor{black}{|{\eta _1}\rangle}  &= \textcolor{black}{\frac{1}{2}(|010\rangle  + |001\rangle  + \sqrt 2 |100\rangle )}\\
\textcolor{black}{|{\eta _2}\rangle}  &= \textcolor{black}{\frac{1}{2}(|010\rangle  + |001\rangle  - \sqrt 2 |100\rangle )}\\
\textcolor{black}{|{\eta _3}\rangle}  &= \textcolor{black}{\frac{1}{2}(|110\rangle  + |101\rangle  + \sqrt 2 |000\rangle )}\\
\textcolor{black}{|{\eta _4}\rangle}  &= \textcolor{black}{\frac{1}{2}(|110\rangle  + |101\rangle  - \sqrt 2 |000\rangle )}
\end{aligned}
\end{equation}

Next, Alice performs a \textcolor{black}{joint} measurement on her \textcolor{black}{three} qubits (qubits 1 and 2 of the shared entangled state, and the input state) \textcolor{black}{under the base in Eq.~\eqref{Eq12}.} \textcolor{black}{The measurement operators for the whole 4-qubit system in Eq.~\eqref{Eq11} are constructed as}
\begin{equation}\label{Eq13}
    \textcolor{black}{{\phi _i} = |{\eta _i}\rangle \langle {\eta _i}| \otimes {I_2},~~~~i = 1,2,3,4}
\end{equation}
And the \textcolor{black}{occurrence probability} of each measurement operator’s outcome is \textcolor{black}{calculated as}
\begin{equation}\label{Eq14}
\begin{aligned}
\textcolor{black}{{P_i^{\rm{W}}}} &= \textcolor{black}{\langle \psi _{{\rm{tot}}}^{E_0^{\rm{W}}}|\phi _i^\dag {\phi _i}|\psi _{{\rm{tot}}}^{E_0^{\rm{W}}}\rangle} \\
 &= \textcolor{black}{\left\{ {\begin{array}{*{20}{c}}
{\frac{{1 - |\beta {|^2}r}}{{4 - 2r}},~~~~i = 1,2}\\
{\frac{{1 - |\alpha {|^2}r}}{{4 - 2r}},~~~~i = 3,4}
\end{array}} \right.}
\end{aligned}
\end{equation}

\textcolor{black}{After Bob receives Alice's measurement result, he has to apply the corresponding weak measurement operators to compensate the effects of the ADC and recover the input state at his end. One weak measurement operator is defined as the inverse of the invertible Kraus operator $e_0$ of the ADC. According to Eq.~\eqref{Eq3}, in the case of ADC, the normalized weak measurement reversal is presented as}
\begin{equation}\label{Eq15}
    \textcolor{black}{m_0 = \begin{bmatrix}
 \sqrt{1 - r} & 0\\
 0 & 1
\end{bmatrix}}
\end{equation}

\textcolor{black}{The weak measurement reversal $m_0$ is from the complete measurement set $\{ {m_0},{{\bar{m}}_0}\}$ with ${{\bar m}_0} = [\sqrt r ,0;0,0]$. In our designed teleportation protocol, we only preserve the result of $m_0$, discard the result of ${{\bar m}_0}$, and normalize the final state at the end of the teleportation process.}

\textcolor{black}{Generally, there is a trade-off between fidelity and success probability in weak measurement-based decoherence control schemes \cite{lab8, lab32}, hence we define the strength of the weak measurement as a variable $q \in [0,1]$, and consider the unitary operators in the last step of the original quantum teleportation protocols. Hence, in the end, the weak measurement reversal operator is defined as} 
\begin{equation}\label{Eq16}
    \textcolor{black}{{M_i} = {U_i}\left[ {\begin{array}{*{20}{c}}
{\sqrt {1 - q} }&0\\
0&1
\end{array}} \right],~~~~i = 1,2,3,4}
\end{equation}
\textcolor{black}{where $U_1 = I_2,~U_2 = \sigma_z,~U_3 =\sigma_x$ and $U_4 = \sigma_x \sigma_z $ with $\sigma_i~(i = x,y,z)$ being Pauli operators, which depends on the result of Alice's joint measurement. }

The weak measurement operators \textcolor{black}{according to different Alice’s measurement results as well as the non-normalized states of Bob’s qubit are given in Table \ref{Tab1}. }

\begin{table*}[t]
\caption{Alice's measurement results and corresponding Bob's weak measurement operators to \textcolor{black}{recover damped states} in TP-EW.}
\centering
\begin{tabular}{|c|c|c|} \hline
Alice's result & \textcolor{black}{non-normalized state of Bob's qubit} & Bob's weak measurement operator \\ \hline
\textcolor{black}{$|{\eta _1}\rangle $} & \textcolor{black}{$|\psi _{{\eta _1}}^{\rm{W}}\rangle  = \frac{1}{{\sqrt {4 - 2r} }}{(\alpha |0\rangle _3} + \beta \sqrt {1 - r} |1{\rangle _3})$} & \textcolor{black}{${M_1} = {U_1}[\sqrt {1 - q} |0\rangle \langle 0| + |1\rangle \langle 1|]$} \\ \hline
\textcolor{black}{$|{\eta _2}\rangle $} & \textcolor{black}{$|\psi _{{\eta _2}}^{\rm{W}}\rangle  = \frac{1}{{\sqrt {4 - 2r} }}{(\alpha |0\rangle _3} - \beta \sqrt {1 - r} |1{\rangle _3})$} & \textcolor{black}{${M_2} = {U_2}[\sqrt {1 - q} |0\rangle \langle 0| + |1\rangle \langle 1|]$} \\ \hline
\textcolor{black}{$|{\eta _3}\rangle $} & \textcolor{black}{$|\psi _{{\eta _3}}^{\rm{W}}\rangle  = \frac{1}{{\sqrt {4 - 2r} }}{(\beta |0\rangle _3} + \alpha \sqrt {1 - r} |1{\rangle _3})$} & \textcolor{black}{${M_3} = {U_3}[\sqrt {1 - q} |0\rangle \langle 0| + |1\rangle \langle 1|]$} \\ \hline
\textcolor{black}{$|{\eta _4}\rangle $} & \textcolor{black}{$|\psi _{{\eta _4}}^{\rm{W}}\rangle  = \frac{1}{{\sqrt {4 - 2r} }}{(\beta |0\rangle _3} - \alpha \sqrt {1 - r} |1{\rangle _3})$} & \textcolor{black}{${M_4} = {U_4}[\sqrt {1 - q} |0\rangle \langle 0| + |1\rangle \langle 1|]$} \\ \hline
\end{tabular}
\label{Tab1}
\end{table*}

Generally, the output states of Bob\textcolor{black}{'s qubit} corresponding to different Alice's measurement outcomes can be described as follows
\begin{equation}\label{Eq17}
\begin{aligned}
\textcolor{black}{\rho _{{M_i}}^{\rm{W}}} &= \textcolor{black}{\frac{{{M_i}|\psi _{{\eta _i}}^{\rm{W}}\rangle \langle \psi _{{\eta _i}}^{\rm{W}}|M_i^\dag }}{{\langle \psi _{{\eta _i}}^{\rm{W}}|M_i^\dag {M_i}|\psi _{{\eta _i}}^{\rm{W}}\rangle }}}\\
 &\buildrel \Delta \over = \left\{ {\begin{array}{*{20}{c}}
\begin{array}{l}
\textcolor{black}{\frac{1}{{(4 - 2r)g_{{M_i}}^{\rm{W}}}}}\left[ {\begin{array}{*{20}{c}}
\textcolor{black}{{|\alpha {|^2}(1 - q)}}\\
\textcolor{black}{{{\alpha ^ * }\beta \sqrt {1 - q} \sqrt {1 - r} }}
\end{array}} \right.\\
\left. ~~~~{\begin{array}{*{20}{c}}
\textcolor{black}{{\alpha {\beta ^ * }\sqrt {1 - q} \sqrt {1 - r} }}\\
\textcolor{black}{{|\beta {|^2}(1 - r)}}
\end{array}} \right],~~~~\textcolor{black}{i = 1,2}
\end{array}\\
\begin{array}{l}
\textcolor{black}{\frac{1}{{(4 - 2r)g_{{M_i}}^{\rm{W}}}}}\left[ {\begin{array}{*{20}{c}}
\textcolor{black}{{|\alpha {|^2}(1 - r)}}\\
\textcolor{black}{{{\alpha ^ * }\beta \sqrt {1 - q} \sqrt {1 - r} }}
\end{array}} \right.\\
\left. ~~~~{\begin{array}{*{20}{c}}
\textcolor{black}{{\alpha {\beta ^ * }\sqrt {1 - q} \sqrt {1 - r} }}\\
\textcolor{black}{{|\beta {|^2}(1 - q)}}
\end{array}} \right],~~~~\textcolor{black}{i = 3,4}
\end{array}
\end{array}} \right.
\end{aligned}
\end{equation}
where \textcolor{black}{$|\psi _{{\eta_i}}^{\rm{W}}\rangle$'s are the non-normalized states of } Bob's \textcolor{black}{qubit} corresponding to different measurement results of Alice, $M_i$\textcolor{black}{'s are} the corresponding weak measurement operators given in Table \ref{Tab1}, and \textcolor{black}{$g_{{M_i}}^{\rm{W}} \buildrel \Delta \over =  \langle \psi _{{\eta _i}}^{\rm{W}}|M_i^\dag {M_i}|\psi _{{\eta _i}}^{\rm{W}}\rangle $ are} the success probability of gaining the state \textcolor{black}{$\rho _{{M_i}}^{\rm{W}}$} as
\begin{equation}
\label{Eq18}
\textcolor{black}{g_{{M_i}}^{\rm{W}} = \left\{ {\begin{array}{*{20}{c}}
{\frac{1}{{4 - 2r}}\left( {|\alpha {|^2}(1 - q) + |\beta {|^2}(1 - r)} \right),i = 1,2}\\
{\frac{1}{{4 - 2r}}\left( {|\alpha {|^2}(1 - r) + |\beta {|^2}(1 - q)} \right),i = 3,4}
\end{array}} \right.}
\end{equation}

Therefore, the total teleportation success probability of TP-EW can be defined as
\begin{equation}\label{Eq19}
g_{{\rm{tot}}}^{{\rm{TP - EW}}} = \textcolor{black}{\sum\limits_{i = 1}^4 {g_{{M_i}}^{\rm{W}}}  = 1 - \frac{q}{{2 - r}}}
\end{equation}

To evaluate the performance of the proposed TP-EW, we also consider the fidelity between input state Eq. \eqref{Eq5} and the output state of TP-EW in Eq. \eqref{Eq17} as 
\begin{equation}\label{Eq20}
\begin{aligned}
\textcolor{black}{{\rm{fid}}_i^{\rm{W}}} &= \textcolor{black}{\langle {\psi _{{\rm{in}}}}|\rho _{{M_i}}^{\rm{W}}|{\psi _{{\rm{in}}}}\rangle} \\
& = \left\{ {\begin{array}{*{20}{c}}
\begin{array}{l}
\textcolor{black}{\frac{{|\beta {|^4}(1 - r) + |\alpha {|^4}(1 - q)}}{{|\alpha {|^2}(1 - q) + |\beta {|^2}(1 - r)}}}\\
 \textcolor{black}{+ \frac{{2|\alpha {|^2}|\beta {|^2}\sqrt {1 - q} \sqrt {1 - r} }}{{|\alpha {|^2}(1 - q) + |\beta {|^2}(1 - r)}},~~~~i = 1,2}
\end{array}\\
\begin{array}{l}
\textcolor{black}{\frac{{|\beta {|^4}(1 - q) + |\alpha {|^4}(1 - r)}}{{|\alpha {|^2}(1 - r) + |\beta {|^2}(1 - q)}}}\\
 \textcolor{black}{+ \frac{{2|\alpha {|^2}|\beta {|^2}\sqrt {1 - q} \sqrt {1 - r} }}{{|\alpha {|^2}(1 - r) + |\beta {|^2}(1 - q)}},~~~~i = 3,4}
\end{array}
\end{array}} \right.
\end{aligned}
\end{equation}

Hence, the average teleportation fidelity of proposed TP-EW over all possible input states is 
\begin{equation}\label{Eq21}
{\rm{Fid}}_{{\rm{av}}}^{{\rm{TP - EW}}} = \textcolor{black}{\int_0^1 {\sum\limits_{i = 1}^4 {P_i^{\rm{W}}{\rm{fid}}_i^{\rm{W}}} {\rm{d}}|\alpha {|^2}} }
\end{equation}

\textcolor{black}{Here, let us examine the effects of the strength of weak measurement on the performance of the proposed TP-EW according to Eqs.~\eqref{Eq19}--\eqref{Eq21}:}

\noindent{1)} \textcolor{black}{When $q = r$, the average teleportation fidelity ${\rm{Fid}}_{{\rm{av}}}^{{\rm{TP - EW}}}$ is equal to 1 with the corresponding total teleportation success probability of $g_{{\rm{tot}}}^{{\rm{TP - EW}}} = 1 - r/(2 - r)$. }

\noindent{2)} \textcolor{black}{When $q = 0$, no weak measurement is applied; hence, the teleportation protocol remains unchanged and the scheme becomes an entanglement distribution process followed by the deterministic standard teleportation. In other words, after a successful EAM during entanglement distribution process, we proceed to the deterministic original teleportation protocol. In other words, after a successful EAM during the entanglement distribution process, we proceed to the deterministic original teleportation protocol. The average teleportation fidelity ${\rm{Fid}}_{{\rm{av}}}^{{\rm{TP - EW}}}$ is always less than 1 with the corresponding total teleportation success probability of unity. }

\noindent{3)} \textcolor{black}{When $q \in (0, r)$, one can strike a balance between the average teleportation fidelity and the success probability, i.e., both ${\rm{Fid}}_{{\rm{av}}}^{{\rm{TP - EW}}}$ and $g_{{\rm{tot}}}^{{\rm{TP - EW}}}$ vary between those in the case of $q = 0$ and those in the case of $q = r$. }

\noindent{4)} \textcolor{black}{However, a value of $q$ within $(r, 1]$ is not acceptable, since both the average teleportation fidelity ${\rm{Fid}}_{{\rm{av}}}^{{\rm{TP - EW}}}$ and the total teleportation success probability $g_{{\rm{tot}}}^{{\rm{TP - EW}}}$ are lower than those in the case of $q = r$. }

For comparison, we consider the \textcolor{black}{original teleportation protocol with the W or Bell state} through an ADC \textcolor{black}{under} no protection. The average teleportation fidelity is calculated as 
\begin{equation}\label{Eq22}
    \textcolor{black}{{\rm{Fid}}_{{\rm{av}}}^{{\rm{W/Bell(ori)}}}} = \frac{1}{{30}}\left( {8\sqrt {1 - r}  + 22 - 7r} \right)
\end{equation}
\textcolor{black}{The detailed procedure for the calculation of the average teleportation fidelity of the original teleportation protocol with the W state through an ADC under no protection is given in Appendix B, and the result on the average teleportation fidelity of the original teleportation protocol with the Bell state are available from Eq.~(A6) of Ref.~\cite{lab10}. }

\textcolor{black}{For further comparison, we consider a probabilistic teleportation protocol which is based on weak measurement, namely, the MR framework of teleportation \cite{lab8}. In the MR framework, Bob applies designed weak measurement reversals instead of unitary operations, to suppress the effects of the ADC. By considering the maximally entangled Bell state as}
\begin{equation}\label{Eq23}
    \textcolor{black}{|\psi {\rangle _{{\rm{ab}}}}{\rm{ = }}\frac{1}{{\sqrt 2 }}\left( {|0{\rangle _{\rm{a}}}|0{\rangle _{\rm{b}}} + |1{\rangle _{\rm{a}}}|1{\rangle _{\rm{b}}}} \right)}
\end{equation}

The average teleportation fidelity of \textcolor{black}{the MR framework through an ADC over all possible input states} is presented as
\begin{equation}\label{Eq24}
\textcolor{black}{{\rm{Fid}}_{{\rm{av}}}^{{\rm{MR}}}=\int_0^1 {\left( {\frac{{1\!+\!r|\alpha {|^2}|\beta {|^2}}}{{2(1\!+\! r|\beta {|^2})}}\!+\!\frac{{1\!+\!r|\alpha {|^2}|\beta {|^2}}}{{2(1\!+\!r|\alpha {|^2})}}} \right){\rm{d}}|\alpha {|^2}} }
\end{equation}

\textcolor{black}{The MR framework of teleportation is probabilistic due to the incompleteness of weak measurement reversal employed in the last step of the teleportation procedure. The total teleportation success probability of the MR framework is }
\begin{equation}\label{Eq25}
    \textcolor{black}{g_{{\rm{tot}}}^{{\rm{MR}}} = \frac{{2 - r - {r^2}}}{2}}
\end{equation}

\textcolor{black}{For fair comparison, we also investigate our proposed TP-EW with the maximally entangled Bell state and prove that all performance indicators, including the average teleportation fidelity and total teleportation success probability, are exactly the same as those of the TP-EW with the W state. The detailed procedure of the TP-EW with the Bell state is given in Appendix C. }

\subsection{\textcolor{black}{Numerical simulation and comparison}}
\begin{figure}
\centering
\subfigure[]
{
\centering
\includegraphics[height=6.5cm]{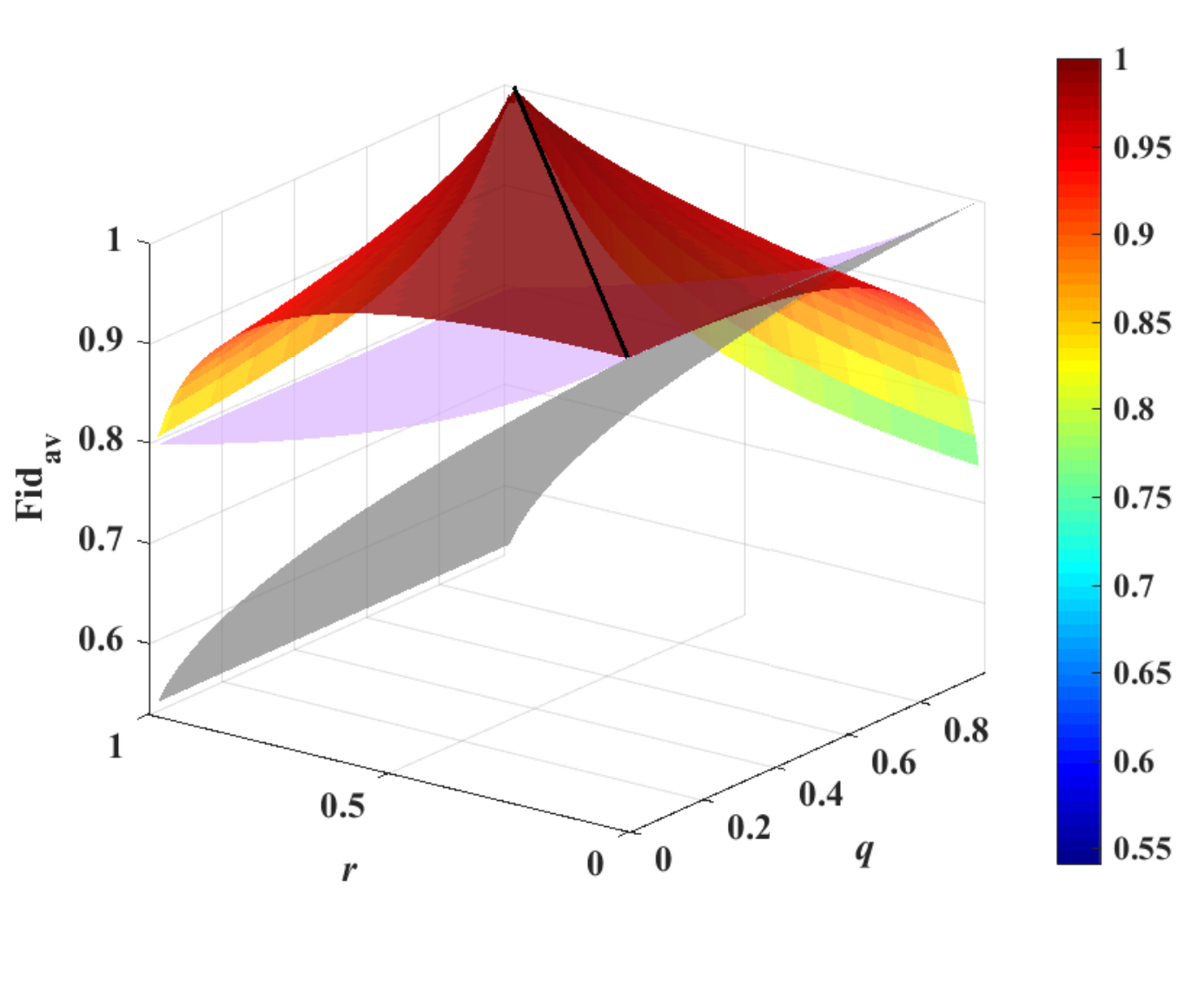}
\label{fig2a}
}%
\\
\subfigure[]
{
\centering
\includegraphics[height=6.5cm]{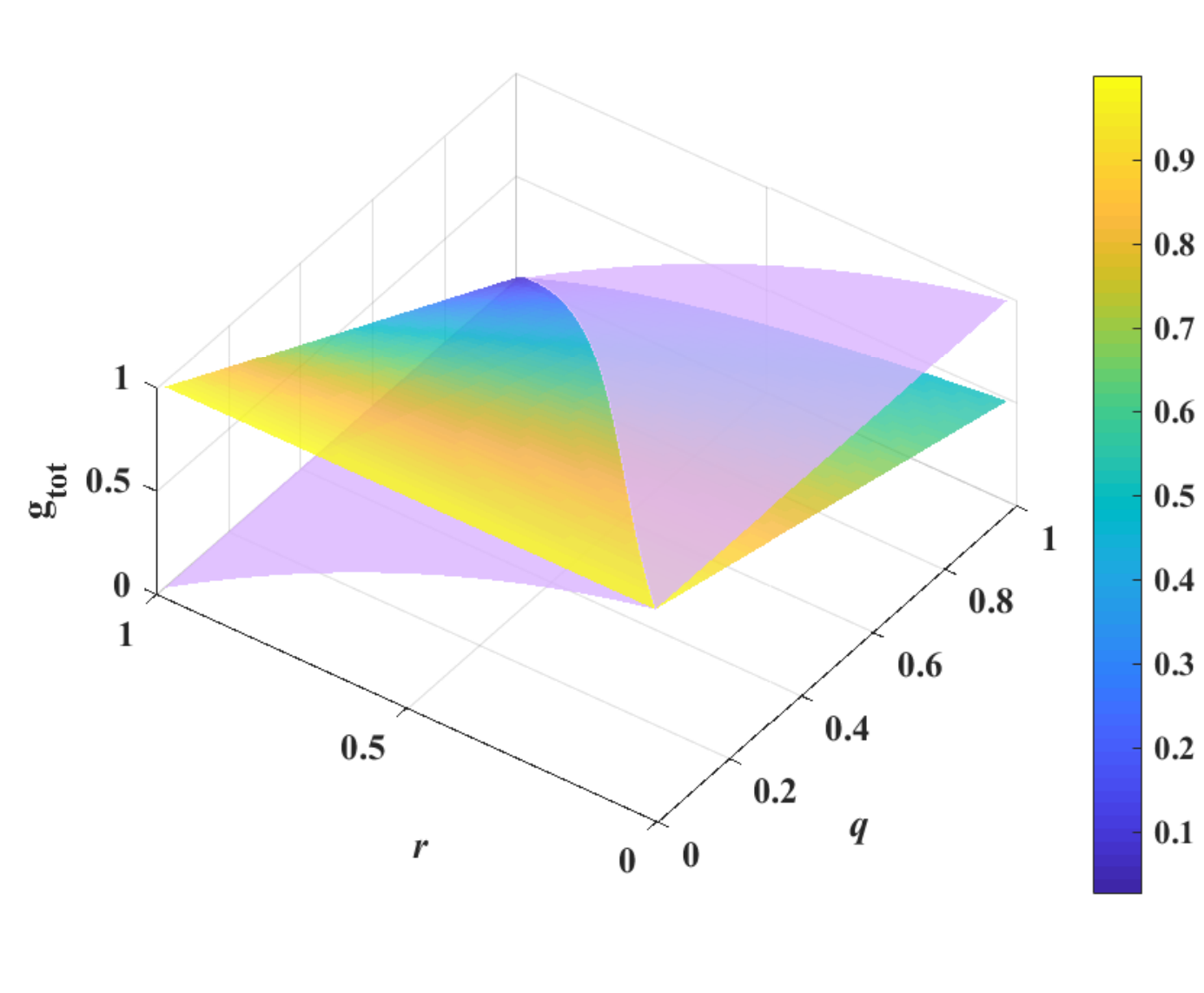}
\label{fig2b}
}%
\caption{(a)~The average teleportation fidelity of TP-EW as a function of decaying rate $r$ and the weak measurement strength $q$. The lavender plate is the average teleportation fidelity of the \textcolor{black}{MR framework}, and the gray plate is the average fidelity of the \textcolor{black}{original} teleportation protocol \textcolor{black}{under} no protection. The black line indicates the maximum average teleportation fidelity of TP-EW. (b)~The total teleportation success probability of TP-EW as a function of decaying rate $r$ and the weak measurement strength $q$. The lavender plate is the total teleportation success probability of the \textcolor{black}{MR framework}.}
\label{fig2}
\end{figure}

In Fig.~\ref{fig2a}, we plot the average teleportation fidelity of TP-EW ${\rm{Fid}}_{{\rm{av}}}^{{\rm{TP - EW}}}$ in Eq.~\eqref{Eq21}, the average teleportation fidelity of the MR framework ${\rm{Fid}}_{{\rm{av}}}^{{\rm{MR}}}$ in Eq.~\eqref{Eq24} (the lavender plate), and the the average teleportation fidelity of the \textcolor{black}{original} teleportation \textcolor{black}{under} no protection in Eq.~\eqref{Eq22} (the gray plate) as a function of the decaying rate $r$ and the weak measurement strength $q$. Moreover, the total teleportation success probability of TP-EW $g_{{\rm{tot}}}^{{\rm{TP - EW}}}$ in Eq.~\eqref{Eq19} and the total teleportation success probability of the MR framework $g_{{\rm{tot}}}^{{\rm{MR}}}$ in Eq.~\eqref{Eq25} (the lavender plate) as a function of decaying rate $r$ and the weak measurement strength $q$ is given in Fig.~\ref{fig2b}. 

As Fig.~\ref{fig2} depicted, the proposed TP-EW significantly improves the average teleportation fidelity compare to \textcolor{black}{the MR framework} (the lavender plate) and the \textcolor{black}{original} teleportation \textcolor{black}{under} no protection (the gray plate). The average teleportation fidelity of TP-EW can be higher than \textcolor{black}{that of the MR framework} for all amounts of decaying rates by choosing \textcolor{black}{an} appropriate weak measurement strength $q$. By contrasting Fig.~\ref{fig2a} with Fig.~\ref{fig2b} \textcolor{black}{under} the same $r$, it can be inferred that a smaller $q$ within the interval $[0, r]$ leads to the gentler improvement of the average teleportation fidelity with less decreasing in total teleportation success probability. Obviously, the maximum average teleportation fidelity of TP-EW is illustrated by the black line which is equal to one \textcolor{black}{in the case of $q = r$; h}owever, as Fig.~\ref{fig2b} illustrates, the total teleportation success probability of our proposed TP-EW is lower than \textcolor{black}{that of the MR framework} for smaller decaying rates. Particularly, in the case of $q=0$(the ridge lines of the TP-EW curves in Figs.~\ref{fig2a} and \ref{fig2b}), the average teleportation fidelity is improved without loss of the total teleportation success probability compared to \textcolor{black}{the MR framework}. 

\begin{figure}
\centering
\subfigure[]
{
\centering
\includegraphics[height=6.5cm]{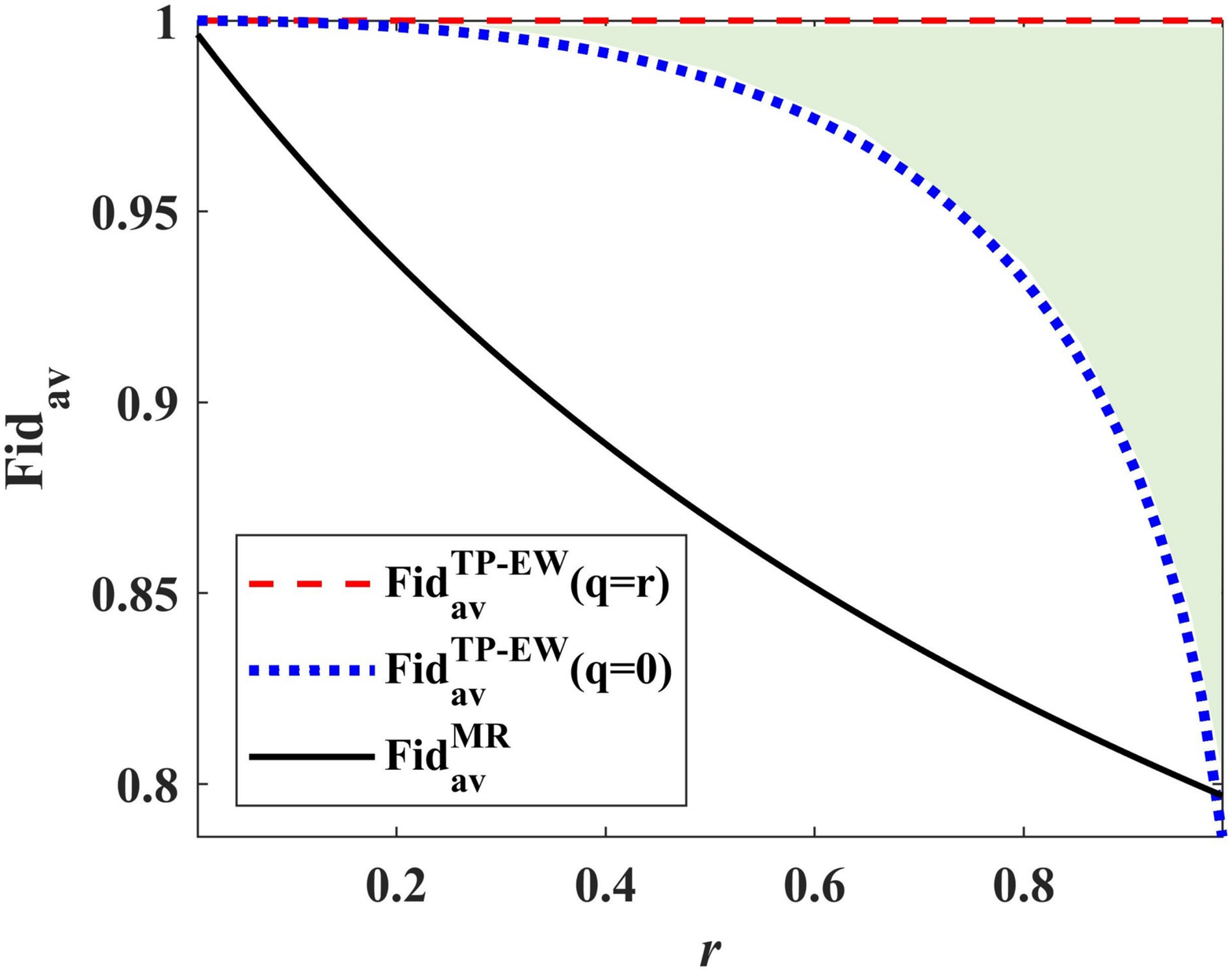}
\label{fig3a}
}%
\\
\subfigure[]
{
\centering
\includegraphics[height=6.5cm]{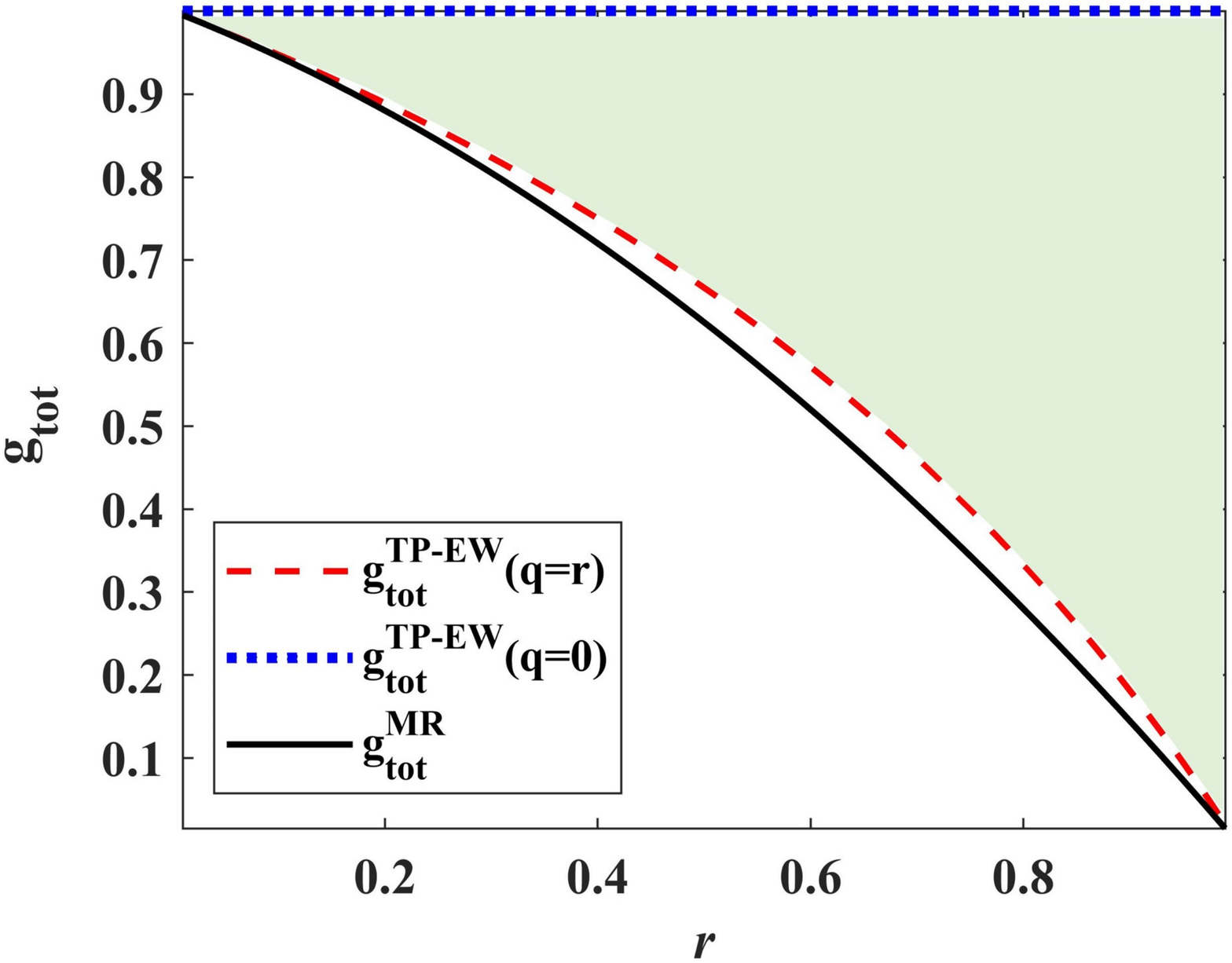}
\label{fig3b}
}%
\caption{(a)~Average teleportation fidelit\textcolor{black}{ies} vs. $r$. (b)~Total teleportation success probabilit\textcolor{black}{ies} vs. $r$. The colored region represents \textcolor{black}{average teleportation fidelities/}total teleportation success probabilities \textcolor{black}{of TP-EW} by varying the weak measurement strength $q$ \textcolor{black}{within $[0,r]$}. }
\label{fig3}
\end{figure}

\textcolor{black}{The c}omparison results of three schemes\textcolor{black}{---the TP-EW with $q = r$ and $q = 0$, and the MR framework---are} shown in Fig.~\ref{fig3}. In Fig.~\ref{fig3a}, we plot 1)~the average teleportation fidelity of TP-EW ${\rm{Fid}}_{{\rm{av}}}^{{\rm{TP - EW}}}$ in \textcolor{black}{Eq.~\eqref{Eq21} with $q = r$} which is the maximized \textcolor{black}{average} fidelity of TP-EW \textcolor{black}{with its }optimum weak measurement strength, 2)~average teleportation fidelity ${\rm{Fid}}_{{\rm{av}}}^{{\rm{TP - EW}}}$ in \textcolor{black}{Eq.~\eqref{Eq21}} by considering EAM \textcolor{black}{but} without applying the weak measurement reversal \textcolor{black}{($q = 0$)}, and 3)~the average teleportation fidelity of \textcolor{black}{the MR framework} ${\rm{Fid}}_{{\rm{av}}}^{{{\rm{\textcolor{black}{MR}}}}}$ in Eq.~\eqref{Eq24}. Moreover, Fig.~\ref{fig3b} is the comparison results of 1)~the corresponding total teleportation success probability of TP-EW $g_{{\rm{tot}}}^{{\rm{TP - EW}}}$ in Eq.~\eqref{Eq19} \textcolor{black}{with $q = r$}, 2)~the total teleportation success probability \textcolor{black}{of TP-EW in Eq.~\eqref{Eq19} with $q = 0$}, and 3)~the total teleportation success probability of \textcolor{black}{the MR framework} in Eq.~\eqref{Eq25}.

By contrasting Fig.~\ref{fig3a} with Fig.~\ref{fig3b}, \textcolor{black}{it is inferred that} the average teleportation fidelity equal to one is achieved \textcolor{black}{by setting $q = r$} even for intense decaying rates with \textcolor{black}{the corresponding} teleportation success probability \textcolor{black}{higher than that of the MR framework}. Also, \textcolor{black}{when $q = 0$} (the blue dotted line), the TP-EW can improve the teleportation fidelity significantly compared to \textcolor{black}{the MR framework} with \textcolor{black}{a total teleportation} success probabilit\textcolor{black}{y of unity} for \textcolor{black}{all} decaying rates. In fact, the colored region is attainable by varying the weak measurement strength $q$ \textcolor{black}{within the interval $[0, r]$}, where the average teleportation fidelity increases and the total success probability decreases \textcolor{black}{as} the weak measurement strength \textcolor{black}{is increased} from 0 to $r$. 

\section{Controlled teleportation through noisy channels by utilizing \textcolor{black}{weak measurement and} environment-assisted measurement}
In this section, we study controlled teleportation protocol\textcolor{black}{s} through noisy channels by utilizing EAM \textcolor{black}{and weak measurement}. Different from Section 2, here we assume that the shared W state is prepared by a third party (Charlie) who delivers the first two qubits to Alice and the third qubit to Bob through \textcolor{black}{independent ADCs}. For simplicity, we assume that the decoherence process occurs for \textcolor{black}{all the qubits of the shared entanglement} locally and independently but with the same decay rate $r$. \textcolor{black}{Both Alice and Bob perform the EAM and tell their results to Charlie. If all ADCs are detected to be in an unexcited state, they are allowed to continue the teleportation process; otherwise, they have to discard the results and restart the entanglement distribution process. At the end of the teleportation process, Bob applies the weak measurement on his qubit if necessary. In the following subsections, we provide the detailed procedure of the proposed controlled teleportation with the W state and Bell state via EAM and the necessary weak measurement.} 

\subsection{Controlled teleportation with W state}
\textcolor{black}{In this subsection, we investigate the controlled teleportation with the W state through ADCs. }

\textcolor{black}{Charlie prepares the W state in Eq.~\eqref{Eq6}, and delivers the first two qubits to Alice and the third qubit to Bob through independent ADCs. The applied Kraus operators for the shared entanglement are}
\begin{equation}\label{Eq26}
    \textcolor{black}{E_i^{{\rm{CW}}} = {e_j} \otimes {e_k} \otimes {e_l}~~~~\text{for}~j,k,l = 0,1}
\end{equation}

\textcolor{black}{Due to the principle of EAM,} we only keep the \textcolor{black}{quantum trajectories} corresponding to \textcolor{black}{the invertible Kraus operator $E_0^{\rm{CW}}$} and discard \textcolor{black}{the quantum trajectories} corresponding to \textcolor{black}{other Kraus operators during the entanglement distribution process}. In this way, the quantum channel between two partners has been successfully constructed and \textcolor{black}{the normalized state of the shared entanglement} can be described as
\begin{equation}\label{Eq27}
\begin{aligned}
\textcolor{black}{|{{\rm{W}}^{E_0^{{\rm{CW}}}}}{\rangle _{123}}} =& \textcolor{black}{E_0^{{\rm{CW}}}|{\rm{W}}{\rangle _{123}}}\\
=& \textcolor{black}{\frac{1}{{2\sqrt {1 - r} }}{(\sqrt {1 - r} |100\rangle _{123}} + \sqrt {1 - r}} \\
& \textcolor{black}{\times |010{\rangle _{123}} + \sqrt 2 \sqrt {1 - r} |001{\rangle _{123}})}\\
 =& \textcolor{black}{\frac{1}{2}{(|100\rangle _{123}} + |010{\rangle _{123}} + \sqrt 2 |001{\rangle _{123}})}
\end{aligned}
\end{equation}

\textcolor{black}{The success probability of entanglement distribution via EAM in controlled teleportation with the W state is calculated as}
\begin{equation}\label{Eq28}
    \textcolor{black}{g_{{\rm{EAM}}}^{{\rm{CW}}} = {}_{123}{\langle {\rm{W}}|{(E_0^{{\rm{CW}}})^\dag }E_0^{{\rm{CW}}}|{\rm{W}}\rangle _{123}} = 1 - r}
\end{equation}

\textcolor{black}{According to Eq.~\eqref{Eq27}, after a successful entanglement distribution via EAM, no decoherence occurs to the W state, and the original teleportation protocol can be started, just as the noise-free case.}

\textcolor{black}{To start the teleportation,} Alice interacts the input qubit with her qubit of the shared entangled state, and the state of the \textcolor{black}{total} system becomes
\begin{equation}\label{Eq29}
\begin{aligned}
\textcolor{black}{|\psi _{{\rm{tot}}}^{E_0^{{\rm{CW}}}}\rangle}  =& (\alpha |0\rangle  + \beta |1\rangle {)_{{\rm{\textcolor{black}{in}}}}} \textcolor{black}{\otimes |{{\rm{W}}^{E_0^{{\rm{CW}}}}}{\rangle _{123}}}\\
 =& \textcolor{black}{|{\eta _1}{\rangle _{{\rm{in,12}}}}{(\alpha |0\rangle _3} + \beta |1{\rangle _3})}\\
 &\textcolor{black}{+ |{\eta _2}{\rangle _{{\rm{in,12}}}}{(\alpha |0\rangle _3} - \beta |1{\rangle _3})}\\
 &\textcolor{black}{+|{\eta _3}{\rangle _{{\rm{in,12}}}}{(\beta |0\rangle _3} + \alpha |1{\rangle _3})}\\
 &\textcolor{black}{+ |{\eta _4}{\rangle _{{\rm{in,12}}}}{(\beta |0\rangle _3} - \alpha |1{\rangle _3})}
\end{aligned}
\end{equation}
where the definition\textcolor{black}{s} of \textcolor{black}{$|\eta_i\rangle $'s} are the same as \textcolor{black}{those in Eq.~\eqref{Eq12}}.

\textcolor{black}{According to Eq.~\eqref{Eq29}, Bob can get the exact input qubit by following the original teleportation and without applying weak measurement. Therefore, in controlled teleportation with the W state, the original teleportation protocol obtains the teleportation fidelity of unity for all possible input qubits after a successful entanglement distribution via EAM. Hence, the average teleportation fidelity of controlled teleportation with the W state via EAM is}
\begin{equation}\label{Eq30}
    \textcolor{black}{{\rm{Fid}}_{{\rm{av - W}}}^{{\rm{CTP - EAM}}} = 1}
\end{equation}

This is because the \textcolor{black}{considered} W state \textcolor{black}{subjected to the amplitude damping noise} is completely symmetric after a successful EAM, which is favorable for quantum communication and computation.

\textcolor{black}{Note that the success probability in Eq.~\eqref{Eq28} is related to entanglement distribution process, and we consider proceeding to the original teleportation protocol with the W state after a successful entanglement distribution. Therefore, the total teleportation success probability of the controlled teleportation with the W state via EAM is also equal to one.}

\textcolor{black}{For comparison, we also consider the controlled teleportation with the W state through ADCs under no protection with the average teleportation fidelity as}
\begin{equation}\label{Eq31}
    \textcolor{black}{{\rm{Fid}}_{{\rm{av}}}^{{\rm{CW(ori)}}} = 1 - \frac{{11}}{{15}}r}
\end{equation}

\subsection{Controlled teleportation with Bell state}
\textcolor{black}{In this subsection, we study the controlled teleportation protocol with the Bell state through ADCs by utilizing EAM and weak measurement.}

\textcolor{black}{Charlie prepares the Bell state in Eq.~\eqref{Eq23}, and delivers the first qubit to Alice and the second qubit to Bob through independent ADCs. Since both qubits of the entangled pair pass through the ADC, the applied Kraus operators become}
\begin{equation}\label{Eq32}
    \textcolor{black}{E_i^{{\rm{CB}}} = {e_j} \otimes {e_k}~~~~\text{for}~j,k = 0,1}
\end{equation}
\textcolor{black}{where only $E_0^{{\rm{CB}}} = {e_0} \otimes {e_0}$ is invertible; hence after Bob applies EAM on the ADC, the quantum trajectories corresponding to $E_0^{{\rm{CB}}}$ are kept, and the quantum trajectories corresponding to other Kraus operators are discarded. }

\textcolor{black}{As a result, after a successful entanglement distribution via EAM, the normalized state of the shared entanglement between Alice and Bob is described as}
\begin{equation}\label{Eq33}
\begin{aligned}
\textcolor{black}{\rho _{{\rm{ab}}}^{E_{\rm{0}}^{{\rm{CB}}}}} &= \textcolor{black}{\frac{{E_{\rm{0}}^{{\rm{CB}}}|\psi {\rangle _{{\rm{ab}}}}{}_{{\rm{ab}}}\langle \psi |{{(E_{\rm{0}}^{{\rm{CB}}})}^\dag }}}{{{}_{{\rm{ab}}}{{\langle \psi |{{(E_{\rm{0}}^{{\rm{CB}}})}^\dag }E_{\rm{0}}^{{\rm{CB}}}|\psi \rangle }_{{\rm{ab}}}}}}}\\
 &\buildrel \Delta \over = \textcolor{black}{\frac{1}{{2g_{{\rm{EAM}}}^{{\rm{CB}}}}}}\left[ {\begin{array}{*{20}{c}}
\textcolor{black}{1}&\textcolor{black}{0}&\textcolor{black}{0}&\textcolor{black}{{1 - r}}\\
\textcolor{black}{0}&\textcolor{black}{0}&\textcolor{black}{0}&\textcolor{black}{0}\\
\textcolor{black}{0}&\textcolor{black}{0}&\textcolor{black}{0}&\textcolor{black}{0}\\
\textcolor{black}{{1 - r}}&\textcolor{black}{0}&\textcolor{black}{0}&\textcolor{black}{{{{(1 - r)}^2}}}
\end{array}} \right]
\end{aligned}
\end{equation}
\textcolor{black}{where $g_{{\rm{EAM}}}^{{\rm{CB}}} \buildrel \Delta \over = {}_{{\rm{ab}}}{\langle \psi |{(E_{\rm{0}}^{{\rm{CB}}})^\dag }E_{\rm{0}}^{{\rm{CB}}}|\psi \rangle _{{\rm{ab}}}}$ is the success probability of entanglement distribution via EAM in controlled teleportation with the Bell state as}
\begin{equation}\label{Eq34}
    \textcolor{black}{g_{{\rm{EAM}}}^{{\rm{CB}}} = \frac{1}{2}\left( {{{\left( {1 - r} \right)}^2} + 1} \right)}
\end{equation}

\textcolor{black}{After the successful entanglement distribution via EAM and employing the shared entangled state in Eq.~\eqref{Eq33}, Alice and Bob proceed to the teleportation protocol via weak measurement. Apart from the last step, detailed steps of the modified quantum teleportation protocol are almost the same as those in Appendix C. However, it is noted that the probabilities of gaining the measurement outcome corresponding to each measurement operator $B_i$ are calculated as}
\begin{equation}\label{Eq35}
\begin{aligned}
\textcolor{black}{P_{{B_i}}^{{\rm{CB}}}} &= \textcolor{black}{{\rm{Tr}}\left( {{B_i}({\rho _{{\rm{in}}}} \otimes \rho _{{\rm{ab}}}^{E_{\rm{0}}^{{\rm{CB}}}})B_i^\dag } \right)}\\
 &= \left\{ {\begin{array}{*{20}{c}}
\textcolor{black}{{\frac{{|\beta {|^2}{r^2} - 2|\beta {|^2}r + 1}}{{2({r^2} - 2r + 2)}},~~~~i = 1,2}}\\
\textcolor{black}{{\frac{{|\alpha {|^2}{r^2} - 2|\alpha {|^2}r + 1}}{{2({r^2} - 2r + 2)}},~~~~i = 3,4}}
\end{array}} \right.
\end{aligned}
\end{equation}
\textcolor{black}{where the definitions of $B_i$’s are the same as those in Eq.~\eqref{EqC-5} of Appendix C.} 

\textcolor{black}{After Alice applies a joint Bell state measurement on her two qubits and sends the measurement outcome to Bob through a classical channel, Bob knows that the non-normalized state of his qubit now is described as}
\begin{equation}\label{Eq36}
\begin{aligned}
\textcolor{black}{\rho _{{B_i}}^{{\rm{CB}}}} &= \textcolor{black}{{{\mathop{\rm Tr}\nolimits} _{{\rm{in,a}}}}[{B_i}({\rho _{{\rm{in}}}} \otimes \rho _{{\rm{ab}}}^{E_{\rm{0}}^{{\rm{CB}}}})B_i^\dag ]}\\
 &= \left\{ {
 \textcolor{black}{ \begin{array}{*{20}{c}}
\begin{array}{l}
\frac{1}{{2({r^2} - 2r + 2)}}\\
 \times \left[ {\begin{array}{*{20}{c}}
{|\alpha {|^2}}&{\alpha {\beta ^ * }(1 - r)}\\
{{\alpha ^ * }\beta (1 - r)}&{|\beta {|^2}{{(1 - r)}^2}}
\end{array}} \right],i = 1
\end{array}\\
\begin{array}{l}
\frac{1}{{2({r^2} - 2r + 2)}}\\
 \times \left[ {\begin{array}{*{20}{c}}
{|\alpha {|^2}}&{\alpha {\beta ^ * }(r - 1)}\\
{{\alpha ^ * }\beta (r - 1)}&{|\beta {|^2}{{(1 - r)}^2}}
\end{array}} \right],i = 2
\end{array}\\
\begin{array}{l}
\frac{1}{{2({r^2} - 2r + 2)}}\\
 \times \left[ {\begin{array}{*{20}{c}}
{|\beta {|^2}}&{\alpha {\beta ^ * }(1 - r)}\\
{{\alpha ^ * }\beta (1 - r)}&{|\alpha {|^2}{{(1 - r)}^2}}
\end{array}} \right],i = 3
\end{array}\\
\begin{array}{l}
\frac{1}{{2({r^2} - 2r + 2)}}\\
 \times \left[ {\begin{array}{*{20}{c}}
{|\beta {|^2}}&{\alpha {\beta ^ * }(r - 1)}\\
{{\alpha ^ * }\beta (r - 1)}&{|\alpha {|^2}{{(1 - r)}^2}}
\end{array}} \right],i = 4
\end{array}
\end{array}}
} \right.
\end{aligned}
\end{equation}

\textcolor{black}{To restore the state of Bob’s qubit, weak measurement operators are applied in the last step of the controlled teleportation protocol as}
\begin{equation}\label{Eq37}
    \textcolor{black}{{\tilde M_i}' = {U_i}\left[ {\begin{array}{*{20}{c}}
{1 - q'}&0\\
0&0
\end{array}} \right],~~~~i = 1,2,3,4}
\end{equation}
\textcolor{black}{where $q' \in [0,r]$ is the acceptable strength of the weak measurement reversal, and the definitions of $U_i$’s are the same as those in Eq.~\eqref{Eq6}.}

\textcolor{black}{After Bob applies the weak measurement reversal in Eq.~\eqref{Eq37}, the output state of his qubit becomes}
\begin{equation}\label{Eq38}
\begin{aligned}
\textcolor{black}{\rho _{{M_i}'}^{{\rm{CB}}}} &= \textcolor{black}{\frac{{{{\tilde M}_i}'\rho _{{B_i}}^{{\rm{CB}}}{{({{\tilde M}_i}')}^\dag }}}{{{\rm{Tr}}\left( {{{\tilde M}_i}'\rho _{{B_i}}^{{\rm{CB}}}{{({{\tilde M}_i}')}^\dag }} \right)}}}\\
 & \buildrel \Delta \over = \left\{ {
 \textcolor{black}{ \begin{array}{*{20}{c}}
\begin{array}{l}
\frac{1}{{2({r^2} - 2r + 2)g_i^{{\rm{CB}}}}}\left[ {\begin{array}{*{20}{c}}
{|\alpha {|^2}{{(1 - q')}^2}}\\
{{\alpha ^ * }\beta (1 - r)(1 - q')}
\end{array}} \right.\\
~~~~~\left. {\begin{array}{*{20}{c}}
{\alpha {\beta ^ * }(1 - r)(1 - q')}\\
{|\beta {|^2}{{(1 - r)}^2}}
\end{array}} \right],~~~~i = 1,2
\end{array}\\
\begin{array}{l}
\frac{1}{{2({r^2} - 2r + 2)g_i^{{\rm{CB}}}}}\left[ {\begin{array}{*{20}{c}}
{|\alpha {|^2}{{(1 - r)}^2}}\\
{{\alpha ^ * }\beta (1 - r)(1 - q')}
\end{array}} \right.\\
~~~~~\left. {\begin{array}{*{20}{c}}
{\alpha {\beta ^ * }(1 - r)(1 - q')}\\
{|\beta {|^2}{{(1 - q')}^2}}
\end{array}} \right],~~~~i = 3,4
\end{array}
\end{array}}
} \right.
\end{aligned}
\end{equation}
\textcolor{black}{where $g_i^{{\rm{CB}}} \buildrel \Delta \over = {\rm{Tr}}\left( {{{\tilde M}_i}'\rho _{{B_i}}^{{\rm{CB}}}{{({{\tilde M}_i}')}^\dag }} \right)$ are the success probability of gaining the state $\rho _{{M_i}'}^{{\rm{CB}}}$ as}
\begin{equation}\label{Eq39}
\textcolor{black}{g_i^{{\rm{CB}}}\!=\!\left\{{\begin{array}{*{20}{c}}
\begin{array}{l}
\frac{1}{{2({r^2} - 2r + 2)}}\\
 \times \left( {|\alpha {|^2}{{(1\!-\!q')}^2}\!+\!|\beta {|^2}{{(1\!-\!r)}^2}} \right),i = 1,2
\end{array}\\
\begin{array}{l}
\frac{1}{{2({r^2} - 2r + 2)}}\\
 \times \left( {|\alpha {|^2}{{(1\!-\!r)}^2}\!+\!|\beta {|^2}{{(1\!-\!q')}^2}} \right),i = 3,4
\end{array}
\end{array}} \right.}
\end{equation}

\textcolor{black}{Therefore, the total teleportation success probability of the controlled TP-EW with the Bell state is} 
\begin{equation}\label{Eq40}
    \textcolor{black}{g_{{\rm{tot - Bell}}}^{{\rm{CTP - EW}}} = \sum\limits_{i = 1}^4 {g_i^{{\mathop{\rm CB}\nolimits} }}  = 1 - \frac{{2q' - {{(q')}^2}}}{{{r^2} - 2r + 2}}}
\end{equation}

\textcolor{black}{In the controlled TP-EW with the Bell state, the fidelity between the input state in Eq.~\eqref{Eq5} and the output state in Eq.~\eqref{Eq38} is calculated as}
\begin{equation}\label{Eq41}
\begin{aligned}
\textcolor{black}{{\rm{fid}}_i^{{\rm{CB}}}} &= \textcolor{black}{\langle {\psi _{{\rm{in}}}}|\rho _{{M_i}'}^{{\rm{CB}}}|{\psi _{{\rm{in}}}}\rangle} \\
 &= \left\{ {
 \textcolor{black}{ \begin{array}{*{20}{c}}
{\frac{{{{(|\alpha {|^2}q' + |\beta {|^2}r - 1)}^2}}}{{|\alpha {|^2}(q{'^2} - 2q') + |\beta {|^2}({r^2} - 2r) + 1}},~~~~i = 1,2}\\
{\frac{{{{(|\beta {|^2}q' + |\alpha {|^2}r - 1)}^2}}}{{|\alpha {|^2}({r^2} - 2r) + |\beta {|^2}(q{'^2} - 2q') + 1}},~~~~i = 3,4}
\end{array}}
} \right.
\end{aligned}
\end{equation}

\textcolor{black}{Thus, the average teleportation fidelity of the controlled TP-EW with the Bell state over all possible input states is }
\begin{equation}\label{Eq42}
    \textcolor{black}{{\rm{Fid}}_{{\rm{av - Bell}}}^{{\rm{CTP - EW}}} = \int_0^1 {\sum\limits_{i = 1}^4 {(P_{{B_i}}^{{\rm{CB}}}{\rm{fid}}_i^{{\rm{CB}}})} {\rm{d|}}\alpha {{\rm{|}}^2}}} 
\end{equation}

\textcolor{black}{For comparison, we consider the original controlled teleportation protocol with the Bell state under no protection with the average teleportation fidelity as }
\begin{equation}\label{Eq43}
    \textcolor{black}{{\rm{Fid}}_{{\rm{av}}}^{{\rm{CB(ori)}}} = 1 - \frac{{11}}{{15}}r + \frac{7}{{15}}{r^2}}
\end{equation}

\subsection{Numerical simulation and comparison}
\begin{figure}
\centering
\subfigure[]
{
\centering
\includegraphics[height=6.5cm]{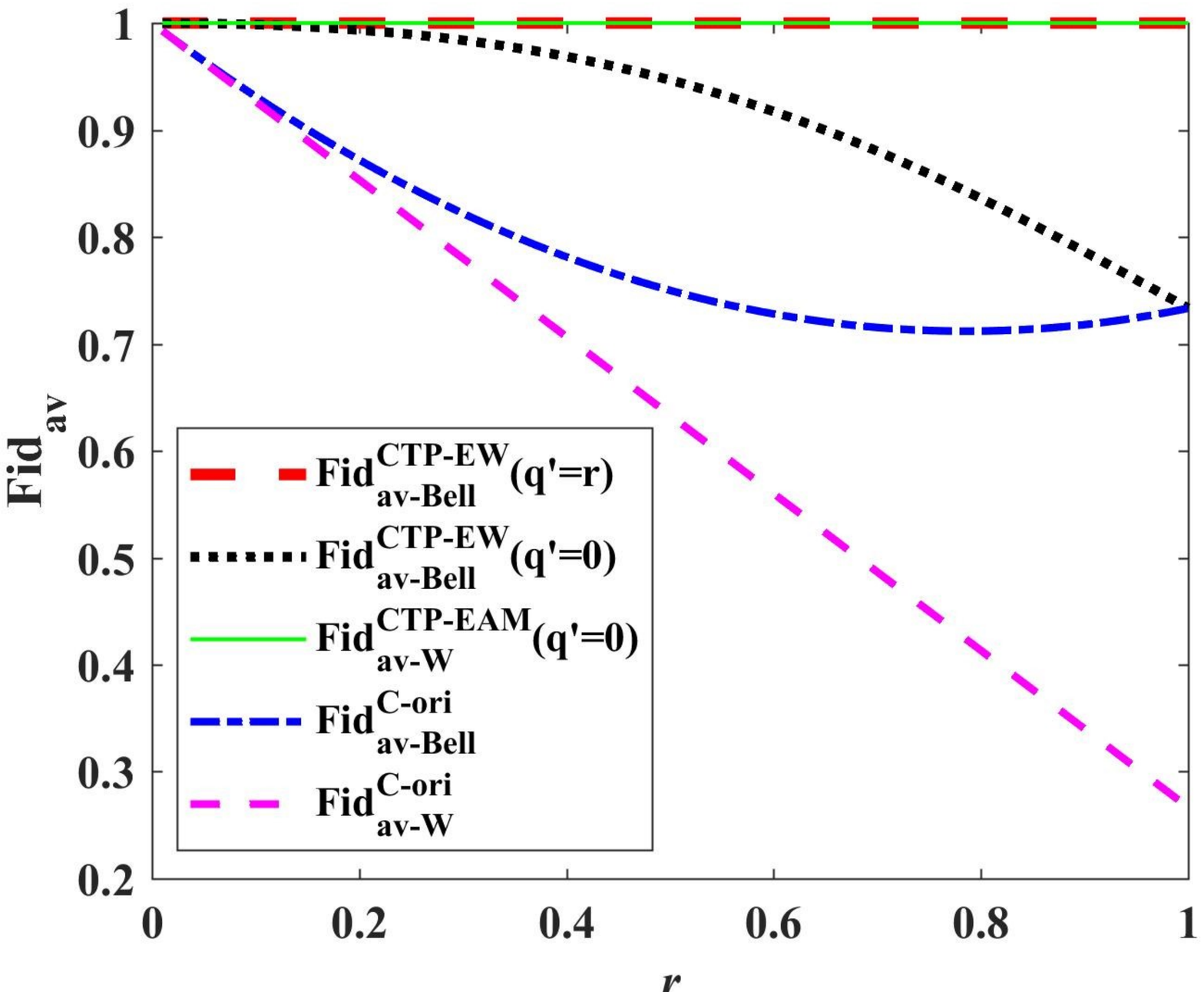}
\label{fig4a}
}%
\\
\subfigure[]
{
\centering
\includegraphics[height=6.5cm]{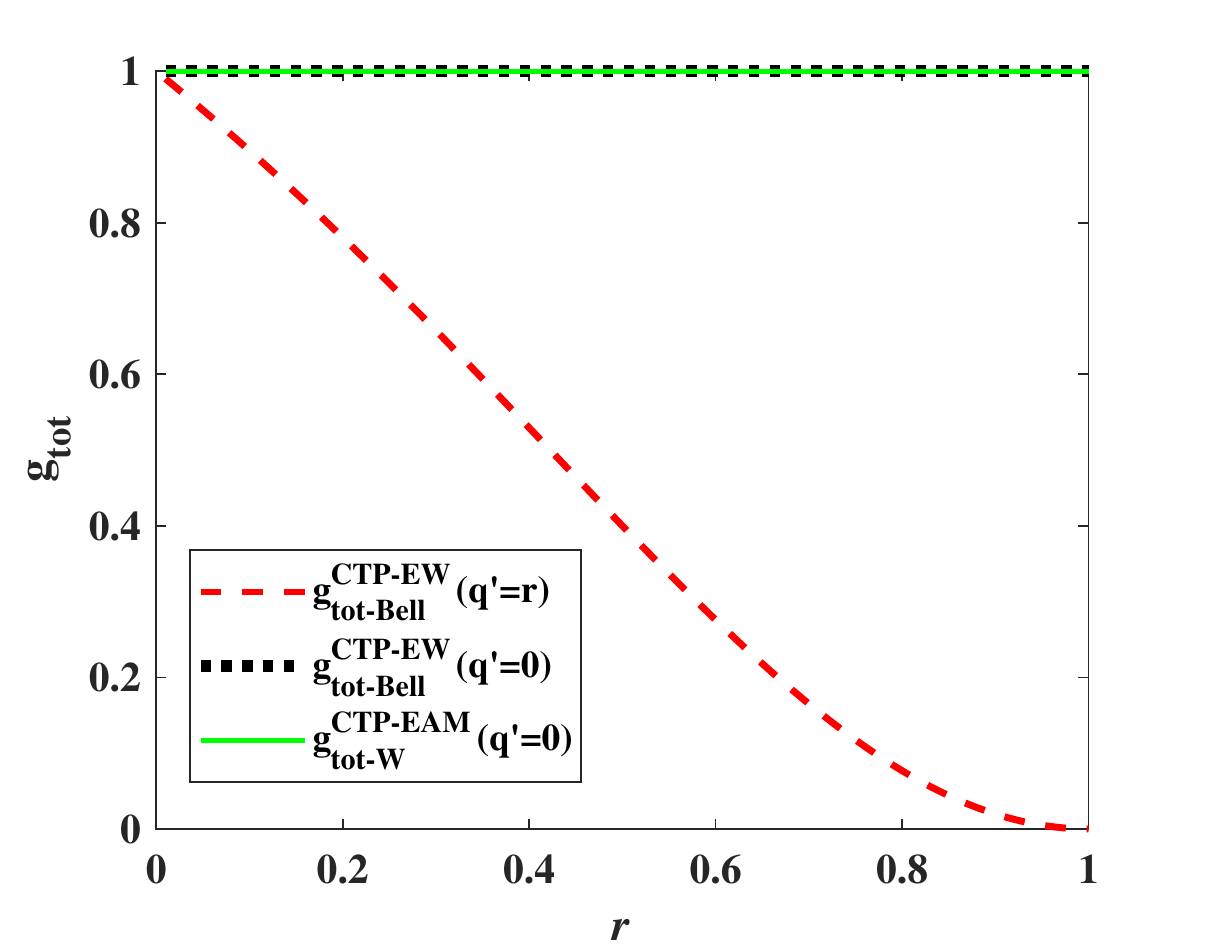}
\label{fig4b}
}%
\caption{\textcolor{black}{(a)~Average teleportation fidelities vs. $r$.} \textcolor{black}{(b)~Total teleportation success probabilities vs. $r$.}}
\label{fig4}
\end{figure}

\textcolor{black}{In this subsection, we compare the average teleportation fidelities and total teleportation success probabilities of the modified controlled teleportation with W and Bell states to the original teleportation protocols under no protection. In Fig.~\ref{fig4a}, we plot 1)~the average teleportation fidelity of the modified controlled teleportation with the W state ${\rm{Fid}}_{{\rm{av - W}}}^{{\rm{CTP - EAM}}}$ in Eq.~\eqref{Eq30}, 2)~the average teleportation fidelity of the controlled TP-EW with the Bell state ${\rm{Fid}}_{{\rm{av - Bell}}}^{{\rm{CTP - EW}}}$ in Eq.~\eqref{Eq42} with $q' = 0$ and $q' = r$, and 3)~the average teleportation fidelity of original controlled teleportation protocols with the W state ${\rm{Fid}}_{{\rm{av}}}^{{\rm{CW(ori)}}}$ in Eq.~\eqref{Eq31} and with the Bell state ${\rm{Fid}}_{{\rm{av}}}^{{\rm{CB(ori)}}}$ in Eq.~\eqref{Eq43} under no protection. Furthermore, in Fig.~\ref{fig4b}, we plot 1)~the total teleportation success probability of controlled teleportation with the W state $g_{{\rm{tot - W}}}^{{\rm{CTP - EAM}}} = 1$, and 2)~the total teleportation success probability of controlled TP-EW with the Bell state $g_{{\rm{tot - Bell}}}^{{\rm{CTP - EW}}}$ in Eq.~\eqref{Eq40} with $q'=0$ and $q' = r$. }

\textcolor{black}{Fig.~\ref{fig4} demonstrates that the modified controlled teleportation with the W state achieves the best performance, i.e., both the average teleportation fidelity and the total teleportation success probability are equal to one. The controlled TP-EW with the Bell state achieves the teleportation fidelity of unity when $q' = r$ at the price of less total teleportation success probabilities in heavy damping cases. Also, when $q'=0$, the average teleportation fidelities of the modified controlled teleportation are remarkably improved compared to the original protocols under no protection, and the corresponding total teleportation success probabilities are always equal to one. Additionally, it can be seen that the original controlled teleportation with the Bell state performs better than that with the W state in the absence of protection. }

\section{Conclusion}
We proposed a \textcolor{black}{high-fidelity }teleportation protocol via EAM and weak measurement through noisy channels \textcolor{black}{with a single copy of the entangled state. The proposed protocol consists of two parts: entanglement distribution via EAM, followed by a modified teleportation protocol by applying designed weak measurement operators in the last step.} The EAM is applied in entanglement distribution step to collect system states corresponding to invertible Kraus operators of the noisy channel. Afterwards, we design\textcolor{black}{ed} weak measurement operators to be applied in the last step of teleportation to reverse the effects of the noise and obtain\textcolor{black}{ed} the \textcolor{black}{average} teleportation fidelity equal to one. \textcolor{black}{The proposed teleportation protocol is applicable to any type of entangled state, but} we \textcolor{black}{only} derived the final expression of the average teleportation fidelity and total teleportation success probability \textcolor{black}{by considering W and Bell entangled states}. Numerical simulation results demonstrated the significant \textcolor{black}{performance} improvement of our proposed TP-EW in comparison with the \textcolor{black}{original} teleportation \textcolor{black}{protocols under} no protection \textcolor{black}{and the MR framework of teleportation}. \textcolor{black}{Furthermore, we investigated the application of TP-EW to} controlled teleportation, \textcolor{black}{in which} all qubits in \textcolor{black}{the shared entanglement pass through independent noisy channels with the same decay rate. The results revealed that by considering controlled teleportation with the W state and just applying EAM during entanglement distribution, the optimal average teleportation fidelity of unity is attained without the need for weak measurement. In addition, by presenting a controlled teleportation protocol with the Bell state using both EAM and weak measurement, we achieved a significant improvement of the average teleportation fidelity in comparison with the original controlled teleportation under no protection.} These results will contribute to the distribution of multi-qubit entanglement in noisy channels and the protection of quantum communication.

\section*{Acknowledgments}
This work was supported by the National Natural Science Foundation of China under Grants 61973290 and 61720106009, and Ministry of Science and Technology of P. R. China Program under the Grant QN2022200007L. 

\section*{\textcolor{black}{Appendix A: Optimal Kraus decomposition of ADC}}
\textcolor{black}{Only one Kraus operator of the ADC is invertible, which means that only one trajectory can be retrieved to the initial state after applying EAM on the ADC. Hence, it is worthwhile to explore how to select the Kraus decomposition to optimize the performance of the proposed TP-EW. This optimization problem can be quite challenging in general. However, in the following, we show how to simplify the problem to find the optimal Kraus decomposition of the ADC.}

\textcolor{black}{For each invertible Kraus operator $e_i$, we can define the weak measurement reversal operator as}
\begin{equation}\label{EqA-1}
    \textcolor{black}{{m_i} = {N_i}e_i^{ - 1}}
\end{equation}
\textcolor{black}{where ${N_i} = \min \left\{ {\sqrt {{\lambda _i}} } \right\}$ are the normalization factor with ${{\lambda _i}}$ being the eigenvalues of the matrix ${e_i}e_i^\dag $.} 

\textcolor{black}{By considering an arbitrary unknown state $|\phi \rangle $, we get}
\begin{equation}\label{EqA-2}
    \textcolor{black}{{m_i}{e_i}|\phi \rangle  = {N_i}|\phi \rangle ,~~~~\forall |\phi \rangle }
\end{equation}

\textcolor{black}{Hence, the unknown state $|\phi \rangle $ is faithfully recovered with the success probability as}
\begin{equation}\label{EqA-3}
    \textcolor{black}{{P_{{e_i}}} = \sum\limits_i {{{({N_i})}^2}}}
\end{equation}

\textcolor{black}{According to Eqs.~\eqref{EqA-3} and \eqref{Eq19}, the total teleportation success probability is also determined by the Kraus decomposition. However, according to Theorem 8.2 of Ref.~\cite{lab40}, the Kraus decomposition is not unique, i.e., arbitrary linear combination of Kraus operators in Eq.~\eqref{Eq7} is valid as long as it has the following form: }
\begin{equation}\label{EqA-4}
    \textcolor{black}{{F_i} = \sum\limits_i {{v_{ij}}{e_i}} }
\end{equation}
\textcolor{black}{where $ v_{ij} \in {\bf{C}}^{2 \times 2} $, and $e_i$’s are the Kraus operators of ADC in Eq.~\eqref{Eq7}. }

\textcolor{black}{Generally, an arbitrary $2\times2$ unitary matrix can be described as}
\begin{equation}\label{EqA-5}
    \textcolor{black}{F_{\alpha,\beta,\gamma,\delta} = \left[ {\begin{array}{*{20}{c}}
{{{\rm{e}}^{i(\alpha  - \beta  - \gamma )}}\cos \delta }&{ - {{\rm{e}}^{i(\alpha  - \beta  + \gamma )}}\sin \delta }\\
{{{\rm{e}}^{i(\alpha  + \beta  - \gamma )}}\sin \delta }&{{{\rm{e}}^{i(\alpha  + \beta  + \gamma )}}\cos \delta }
\end{array}} \right]}
\end{equation}

\textcolor{black}{According to Eq.~\eqref{EqA-5}, it is noted that the success probability only depends on the eigenvalues of the matrix $F_i F_i^\dagger$, for instance,} 
\begin{equation}\label{EqA-6}
\begin{aligned}
    \textcolor{black}{{F_1}F_1^\dag}  =& \textcolor{black}{|{v_{11}}{|^2}{e_1}e_1^\dag  + |{v_{22}}{|^2}{e_2}e_2^\dag } \\
    &\textcolor{black}{+ {v_{11}}v_{12}^ * {e_1}e_2^\dag  + {v_{12}}v_{11}^ * {e_2}e_1^\dag }
\end{aligned}
\end{equation}

\textcolor{black}{To find the eigenvalues of $F_1 F_1^\dagger$, we realize that $\left| {{F_1}F_1^\dag  - \lambda {I_2}} \right| = \left( {|{v_{11}}{|^2} + |{v_{22}}{|^2}|r| - \lambda } \right)\left( {|{v_{11}}{|^2}|1 - r| - \lambda } \right) - |{v_{11}}{|^2}|{v_{22}}{|^2}|r(1 - r)|$ is independent of the phase factors $\alpha$, $\beta$ and $\gamma$; hence we only need to consider $\delta$. Next, we only consider the single-parameter transformation matrix as }
\begin{equation}\label{EqA-7}
    \textcolor{black}{{F_\delta } = \left[ {\begin{array}{*{20}{c}}
{\cos \delta }&{\sin \delta }\\
{\sin \delta }&{\cos \delta }
\end{array}} \right]}
\end{equation}

\begin{figure}[h]
    \centering
    \includegraphics[width=0.45\textwidth]{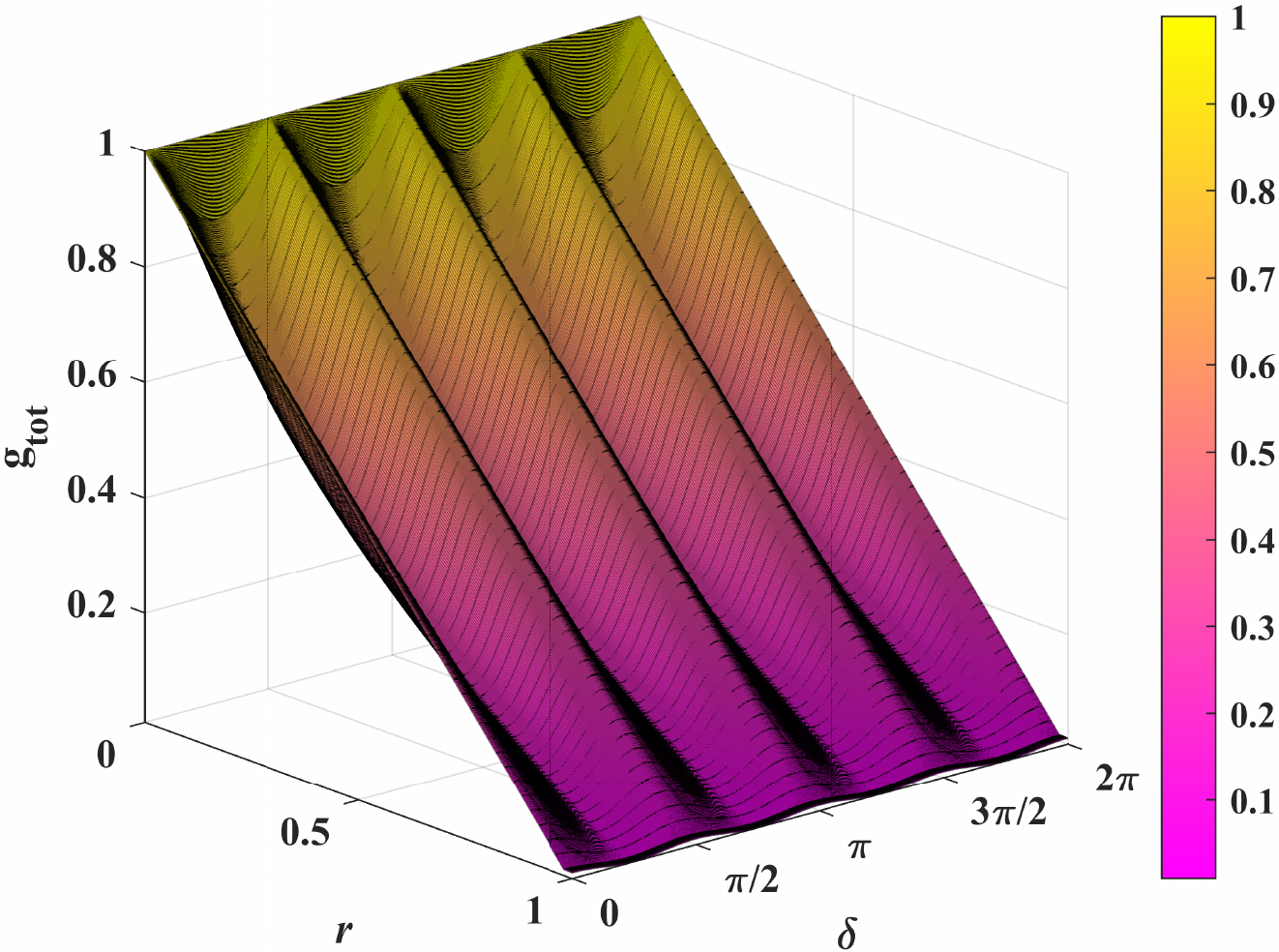}
\caption{\textcolor{black}{Total teleportation success probability for all possible Kraus decompositions.}}
\label{fig5}
\end{figure}

\textcolor{black}{We plot the total teleportation success probability, for different Kraus decompositions according to Eqs.~\eqref{EqA-4} and \eqref{EqA-7}, by varying $r \in [0,1]$ and $\delta \in [0, 2\pi]$ in Fig.~\ref{fig5}. }

\textcolor{black}{As Fig.~\ref{fig5} demonstrates, for all values of $r$, the highest success probability occurs at $\delta = n\pi/2$, for $n \in \bf{Z}$. Thus, it can be concluded that the Kraus decomposition in Eq.~\eqref{Eq7} is the optimal choice to achieve the highest teleportation success probability.}

\section*{\textcolor{black}{Appendix B: Original teleportation protocol with W state through ADC under no protection}}
\textcolor{black}{Here, we study the average teleportation fidelity of the original teleportation protocol with the W state through an ADC under no protection. In this case, the shared W entangled state in Eq.~\eqref{Eq6} after passing through the ADC is }
\begin{equation}\label{EqB-1}
|{{\rm{W}}^{\textcolor{black}{{E_{0,1}^{\rm{W}}}}}}{\rangle _{123}} = E_0^{\rm{\textcolor{black}{W}}}|{\rm{W}}{\rangle _{123}} + E_1^{\rm{\textcolor{black}{W}}}|{\rm{W}}{\rangle _{123}}
\end{equation}
where $E_0^{\rm{\textcolor{black}{W}}}$ and $E_1^{\rm{\textcolor{black}{W}}}$ are the applied Kraus operators given in Eq.~\eqref{Eq8}.

\textcolor{black}{Furthermore,} by considering the original teleportation protocol with the W state and applying unitary operators in the last step, Bob's qubit corresponding to different Alice's measurement outcomes becomes
\begin{equation}\label{EqB-2}
    \rho _{{\textcolor{black}{U_i}}}^{\rm{\textcolor{black}{W}}}\!=\!\left\{ {\begin{array}{*{20}{c}}
{\textcolor{black}{{\frac{1}{{4g_{{U_i}}^{\rm{W}}}}}}\left[ {\begin{array}{*{20}{c}}
{|\alpha {|^2}\!+\!|\beta {|^2}r}&{\alpha {\beta ^ * }\sqrt {1\!-\!r} }\\
{{\alpha ^ * }\beta \sqrt {1\!-\!r} }&{|\beta {|^2}(1\!- \!r)}
\end{array}} \right],i\!=\!1,2}\\
{\textcolor{black}{{\frac{1}{{4g_{{U_i}}^{\rm{W}}}}}}\left[ {\begin{array}{*{20}{c}}
{|\alpha {|^2}(1\!-\!r)}&{\alpha {\beta ^ * }\sqrt {1\!-\!r} }\\
{{\alpha ^ * }\beta \sqrt {1\!-\!r} }&{|\alpha {|^2}r\!+\! |\beta {|^2}}
\end{array}} \right],i\!=\!3,4}
\end{array}} \right.
\end{equation}
\textcolor{black}{where $g_{{U_i}}^{\rm{W}}(i = 1,2,3,4) = 1/4$ are the probabilities of gaining the output state $\rho _{{U_i}}^{\rm{W}}$.}

\textcolor{black}{Also, in order to calculate the average teleportation fidelity, it is necessary to present the probabilities $P_i^{{\rm{W(ori)}}}$'s of gaining measurement outcome corresponding to each measurement operator $\phi_i$ in Eq.~\eqref{Eq13}. Since no result is discarded in the teleportation process, these probabilities are equal to those of gaining the output state $\rho _{{U_i}}^{\rm{W}}$'s, i.e., }
\begin{equation}\label{EqB-3}
    \textcolor{black}{P_i^{{\rm{W(ori)}}} = g_{{U_i}}^{\rm{W}}~~~~{\text{for}}~i=1,2,3,4}
\end{equation}

The teleportation fidelity corresponding to different Alice's measurement results is
\begin{equation}\label{EqB-4}
\begin{aligned}
{\rm{fid}}_i^{{\rm{\textcolor{black}{W(ori)}}}} &= \textcolor{black}{\langle {\psi _{{\rm{in}}}}|\rho _{{U_i}}^{\rm{W}}|{\psi _{{\rm{in}}}}\rangle} \\
 &= \left\{ {\begin{array}{*{20}{c}}
\begin{array}{l}
|\alpha {|^4} + |\beta {|^4}(1 - r)\\
 + |\alpha {|^2}|\beta {|^2}(r + 2\sqrt {1 - r} ),~~i = 1,2
\end{array}\\
\begin{array}{l}
|\beta {|^4} + |\alpha {|^4}(1 - r)\\
 + |\alpha {|^2}|\beta {|^2}(r + 2\sqrt {1 - r} ),~~i = 3,4
\end{array}
\end{array}} \right.
\end{aligned}
\end{equation}

Hence, the average teleportation fidelity of \textcolor{black}{the original} teleportation \textcolor{black}{with the W state} \textcolor{black}{under} no protection through \textcolor{black}{an} ADC is calculated as
\begin{equation}\label{EqB-5}
\begin{aligned}
{\rm{Fid}}_{{\rm{av}}}^{{\rm{\textcolor{black}{W(ori)}}}} &= \textcolor{black}{\int_0^1 {\sum\limits_{i = 1}^4 {P_i^{{\rm{W(ori)}}}{\rm{fid}}_i^{{\rm{W(ori)}}}} {\rm{d}}|\alpha {|^2}}} \\
 &= \frac{1}{{30}}(8\sqrt {1 - r}  + 22 - 7r)
\end{aligned}
\end{equation}

\section*{\textcolor{black}{Appendix C: TP-EW with Bell state through ADC}}
\textcolor{black}{Since only the second qubit of the Bell state passes through the noisy channel, the applied Kraus operators for the whole 2-qubit shared entangled state are }
\begin{equation}\label{EqC-1}
    \textcolor{black}{E_0^{{\rm{Bell}}} = {I_2} \otimes {e_0},~E_1^{{\rm{Bell}}} = {I_2} \otimes {e_1}}
\end{equation}
\textcolor{black}{where $e_i$'s are the Kraus operators of the ADC in Eq.~\eqref{Eq7}. }

\textcolor{black}{Bob now applies EAM to the ADC and informs Alice of the result. If the result of EAM is corresponding to the invertible Kraus operator $E_0^{{\rm{Bell}}}$, the quantum channel between Alice and Bob is successfully established and defined as }
\begin{equation}\label{EqC-2}
\begin{aligned}
\textcolor{black}{\rho _{{\rm{ab}}}^{E_0^{{\rm{Bell}}}}} &= \textcolor{black}{\frac{{E_0^{{\rm{Bell}}}|\psi {\rangle _{{\rm{ab}}}}{}_{{\rm{ab}}}\langle \psi |{{(E_0^{{\rm{Bell}}})}^\dag }}}{{{}_{{\rm{ab}}}{{\langle \psi |{{(E_0^{{\rm{Bell}}})}^\dag }E_0^{{\rm{Bell}}}|\psi \rangle }_{{\rm{ab}}}}}}}\\
 &\buildrel \Delta \over = \textcolor{black}{\frac{1}{{2g_{{\rm{EAM}}}^{{\rm{Bell}}}}}}\left[ {\textcolor{black}{\begin{array}{*{20}{c}}
1&0&0&{\sqrt {1 - r} }\\
0&0&0&0\\
0&0&0&0\\
{\sqrt {1 - r} }&0&0&{1 - r}
\end{array}}} \right]
\end{aligned}
\end{equation}
\textcolor{black}{where $g_{{\rm{EAM}}}^{{\rm{Bell}}} \buildrel \Delta \over = {}_{{\rm{ab}}}{\langle \psi |{(E_0^{{\rm{Bell}}})^\dag }E_0^{{\rm{Bell}}}|\psi \rangle _{{\rm{ab}}}} = 1 - r/2$ denotes the success probability of entanglement distribution before TP-EW with the Bell state, and its value is equal to that before TP-EW with the W state in Eq.~\eqref{Eq10}. }

\textcolor{black}{Alice interacts the input qubit with her half of the entangled pair. Thus, the state of the whole 3-qubit system consisting of the input qubit and the whole shared entanglement is described as}
\begin{equation}\label{EqC-3}
\textcolor{black}{\rho _{{\rm{tot}}}^{E_0^{{\rm{Bell}}}} = {\rho _{{\rm{in}}}} \otimes \rho _{{\rm{ab}}}^{E_0^{{\rm{Bell}}}}}
\end{equation}
\textcolor{black}{where ${\rho _{{\rm{in}}}} = |{\psi _{{\rm{in}}}}\rangle \langle {\psi _{{\rm{in}}}}|$ is the density matrix of the input state $|{\psi _{{\rm{in}}}}\rangle $ in Eq.~\eqref{Eq5}. }

\textcolor{black}{Then, she makes a joint Bell state measurement on her two qubits (the input state and her share of the entangled state) with measurement operators $|{b_i}\rangle \langle {b_i}|$, where $b_i$'s are defined as}
\begin{equation}\label{EqC-4}
\begin{array}{l}
\textcolor{black}{|{b_1}\rangle  = \frac{1}{{\sqrt 2 }}(|00\rangle  + |11\rangle ),~|{b_2}\rangle  = \frac{1}{{\sqrt 2 }}(|00\rangle  - |11\rangle )}\\
\textcolor{black}{|{b_3}\rangle  = \frac{1}{{\sqrt 2 }}(|01\rangle  + |10\rangle ),~|{b_4}\rangle  = \frac{1}{{\sqrt 2 }}(|01\rangle  - |10\rangle )}
\end{array}
\end{equation}
\textcolor{black}{and the measurement operators applied on the whole 3-qubit system in Eq.~\eqref{EqC-3} are constructed as}
\begin{equation}\label{EqC-5}
    \textcolor{black}{{B_i} = |{b_i}\rangle \langle {b_i}| \otimes {I_2},~~~~i = 1,2,3,4}
\end{equation}

\textcolor{black}{After applying the joint measurement, Alice sends the measurement result to Bob through a classical channel. Thus, Bob knows that the non-normalized state of his qubit is described as}
\begin{equation}\label{EqC-6}
    \textcolor{black}{\rho _{{b_i}}^{{\rm{Bell}}} = {\rm{T}}{{\rm{r}}_{{\rm{in,a}}}}\left( {{B_i}\rho _{{\rm{ab}}}^{E_0^{{\rm{Bell}}}}B_i^\dag } \right),~~~~i = 1,2,3,4}
\end{equation}
\textcolor{black}{where ${\rm{T}}{{\rm{r}}_{{\rm{in,a}}}}( \bullet )$ denotes the partial trace over the input qubit and the first qubit of the shared entangled pair, and the occurrence probability of each measurement operator's outcome is calculated as}
\begin{equation}\label{EqC-7}
\begin{aligned}
\textcolor{black}{P_i^{{\rm{Bell}}}} &= \textcolor{black}{{\rm{Tr}}\left( {{B_i}\rho _{{\rm{tot}}}^{E_0^{{\rm{Bell}}}}B_i^\dag } \right)}\\
 &= \left\{ {\begin{array}{*{20}{c}}
{\textcolor{black}{\frac{{1 - |\beta {|^2}r}}{{4 - 2r}},~~~~i = 1,2}}\\
{\textcolor{black}{\frac{{1 - |\alpha {|^2}r}}{{4 - 2r}},~~~~i = 3,4}}
\end{array}} \right.
\end{aligned}
\end{equation}

\textcolor{black}{Finally, as is shown in Table \ref{TabC-1}, Bob applies the corresponding weak measurement operators to his qubit according to Alice’s measurement outcomes, where the definition of $M_i$'s is the same as that in Table \ref{Tab1}. }

\begin{table*}[t]
\caption{\textcolor{black}{Alice's measurement results and corresponding Bob's weak measurement operators to recover damped states in TP-EW with the Bell state.}}
\centering
\begin{tabular}{|c|c|c|} \hline
\textcolor{black}{Alice's result} & \textcolor{black}{non-normalized state of Bob's qubit} & \textcolor{black}{Bob's weak measurement operator} \\ \hline
\textcolor{black}{$|{b_1}\rangle $} & \textcolor{black}{$|\psi _{{b_1}}^{\rm{Bell}}\rangle  = \frac{1}{{\sqrt {4 - 2r} }}{(\alpha |0\rangle _3} + \beta \sqrt {1 - r} |1{\rangle _3})$} & \textcolor{black}{${M_1} = {U_1}[\sqrt {1 - q} |0\rangle \langle 0| + |1\rangle \langle 1|]$} \\ \hline
\textcolor{black}{$|{b_2}\rangle $} & \textcolor{black}{$|\psi _{{b_2}}^{\rm{Bell}}\rangle  = \frac{1}{{\sqrt {4 - 2r} }}{(\alpha |0\rangle _3} - \beta \sqrt {1 - r} |1{\rangle _3})$} & \textcolor{black}{${M_2} = {U_2}[\sqrt {1 - q} |0\rangle \langle 0| + |1\rangle \langle 1|]$} \\ \hline
\textcolor{black}{$|{b_3}\rangle $} & \textcolor{black}{$|\psi _{{b_3}}^{\rm{Bell}}\rangle  = \frac{1}{{\sqrt {4 - 2r} }}{(\beta |0\rangle _3} + \alpha \sqrt {1 - r} |1{\rangle _3})$} & \textcolor{black}{${M_3} = {U_3}[\sqrt {1 - q} |0\rangle \langle 0| + |1\rangle \langle 1|]$} \\ \hline
\textcolor{black}{$|{b_4}\rangle $} & \textcolor{black}{$|\psi _{{b_4}}^{\rm{Bell}}\rangle  = \frac{1}{{\sqrt {4 - 2r} }}{(\beta |0\rangle _3} - \alpha \sqrt {1 - r} |1{\rangle _3})$} & \textcolor{black}{${M_4} = {U_4}[\sqrt {1 - q} |0\rangle \langle 0| + |1\rangle \langle 1|]$} \\ \hline
\end{tabular}
\label{TabC-1}
\end{table*}

\textcolor{black}{Therefore, after applying WM operators in Table \ref{TabC-1}, the output states of Bob’s qubit corresponding to different measurement outcomes of Alice are calculated as}
\begin{equation}\label{EqC-8}
\begin{aligned}
&\textcolor{black}{\rho _{{M_i}}^{{\rm{Bell}}}} = \textcolor{black}{\frac{{{M_i}|\psi _{{b_i}}^{{\rm{Bell}}}\rangle \langle \psi _{{b_i}}^{{\rm{Bell}}}|M_i^\dag }}{{\langle \psi _{{b_i}}^{{\rm{Bell}}}|M_i^\dag {M_i}|\psi _{{b_i}}^{{\rm{Bell}}}\rangle }}}\\
& = \left\{ {\begin{array}{*{20}{c}}
\begin{array}{l}
\left[ {\begin{array}{*{20}{c}}
{\textcolor{black}{|\alpha {|^2}(1 - q)}}&\textcolor{black}{{\alpha {\beta ^ * }\sqrt {1 - q} \sqrt {1 - r} }}\\
\textcolor{black}{{{\alpha ^ * }\beta \sqrt {1 - q} \sqrt {1 - r} }}&\textcolor{black}{{|\beta {|^2}(1 - r)}}
\end{array}} \right],\\
~~~~~~~~~~~~~~~~~~~~~~~~~~~~~~~~~~~~~~~~~~~~~\textcolor{black}{i = 1,2}
\end{array}\\
\begin{array}{l}
\left[ {\begin{array}{*{20}{c}}
{\textcolor{black}{|\alpha {|^2}(1 - r)}}&\textcolor{black}{{\alpha {\beta ^ * }\sqrt {1 - q} \sqrt {1 - r} }}\\
\textcolor{black}{{{\alpha ^ * }\beta \sqrt {1 - q} \sqrt {1 - r} }}&{\textcolor{black}{|\beta {|^2}(1 - q)}}
\end{array}} \right],\\
~~~~~~~~~~~~~~~~~~~~~~~~~~~~~~~~~~~~~~~~~~~~~\textcolor{black}{i = 3,4}
\end{array}
\end{array}} \right.
\end{aligned}
\end{equation}
\textcolor{black}{where ${|\psi _{{b_i}}^{{\rm{Bell}}}\rangle }$'s are the non-normalized states of Bob’s qubit corresponding to different measurement results of Alice, $M_i$'s are the corresponding weak measurement operators given in Table~\ref{TabC-1},  and $g_{{M_i}}^{\rm{Bell}} \buildrel \Delta \over =  \langle \psi _{{b_i}}^{\rm{Bell}}|M_i^\dagger {M_i}|\psi _{{b_i}}^{\rm{Bell}}\rangle$ are the success probability of gaining the state $\rho _{{M_i}}^{{\rm{Bell}}}$ as}
\begin{equation}
\label{EqC-9}
\textcolor{black}{g_{{M_i}}^{\rm{Bell}} = \left\{ {\begin{array}{*{20}{c}}
{\frac{1}{{4 - 2r}}\left( {|\alpha {|^2}(1 - q) + |\beta {|^2}(1 - r)} \right),i = 1,2}\\
{\frac{1}{{4 - 2r}}\left( {|\alpha {|^2}(1 - r) + |\beta {|^2}(1 - q)} \right),i = 3,4}
\end{array}} \right.}
\end{equation}

\textcolor{black}{Therefore, the total teleportation success probability of TP-EW with the Bell state can be defined as}
\begin{equation}\label{EqC-10}
\textcolor{black}{g_{{\rm{tot}}}^{{\rm{TP - EW}}} = \sum\limits_{i = 1}^4 {g_{{M_i}}^{\rm{Bell}}}  = 1 - \frac{q}{{2 - r}}}
\end{equation}

\textcolor{black}{Moreover, the fidelity between input state Eq.~\eqref{Eq5} and the output state of TP-EW with the Bell state in Eq.~\eqref{EqC-8} is}
\begin{equation}\label{EqC-11}
\begin{aligned}
\textcolor{black}{{\rm{fid}}_i^{\rm{Bell}}} &= \textcolor{black}{\langle {\psi _{{\rm{in}}}}|\rho _{{M_i}}^{\rm{Bell}}|{\psi _{{\rm{in}}}}\rangle} \\
& = \left\{ {\begin{array}{*{20}{c}}
\begin{array}{l}
\textcolor{black}{\frac{{|\beta {|^4}(1 - r) + |\alpha {|^4}(1 - q)}}{{|\alpha {|^2}(1 - q) + |\beta {|^2}(1 - r)}}}\\
 \textcolor{black}{+ \frac{{2|\alpha {|^2}|\beta {|^2}\sqrt {1 - q} \sqrt {1 - r} }}{{|\alpha {|^2}(1 - q) + |\beta {|^2}(1 - r)}},~~~~i = 1,2}
\end{array}\\
\begin{array}{l}
\textcolor{black}{\frac{{|\beta {|^4}(1 - q) + |\alpha {|^4}(1 - r)}}{{|\alpha {|^2}(1 - r) + |\beta {|^2}(1 - q)}}}\\
 \textcolor{black}{+ \frac{{2|\alpha {|^2}|\beta {|^2}\sqrt {1 - q} \sqrt {1 - r} }}{{|\alpha {|^2}(1 - r) + |\beta {|^2}(1 - q)}},~~~~i = 3,4}
\end{array}
\end{array}} \right.
\end{aligned}
\end{equation}

\textcolor{black}{Therefore, the average teleportation fidelity of TP-EW with the Bell state over all possible input states is defined as}
\begin{equation}\label{EqC-12}
\textcolor{black}{{\rm{Fid}}_{{\rm{av}}}^{{\rm{TP - EW}}} =} \textcolor{black}{\int_0^1 {\sum\limits_{i = 1}^4 {P_i^{\rm{Bell}}{\rm{fid}}_i^{\rm{Bell}}} {\rm{d}}|\alpha {|^2}} }
\end{equation}

\textcolor{black}{It can be found that all performance indicators in TP-EW with the Bell state are exactly the same as those with the W state, by comparing Eqs.~\eqref{EqC-9}--\eqref{EqC-12} with Eqs.~\eqref{Eq18}--\eqref{Eq21}. }


\begin{thebibliography}{50}%
\bibitem{lab1} Pirandola S, Eisert J, Weedbrook C, Furusawa A and Braunstein S L 2015 \textit{Nat. Photonics} \textbf{9}(10): 641--652 DOI: 10.1038/NPHOTON.2015.154

\bibitem{lab2} Duan L M, Lukin M D, Cirac J I and Zoller P 2001 \textit{Nature} \textbf{414}(62): 413--418 DOI: 10.1038/35106500

\bibitem{lab3} Bennett C H, Brassard G, Crepeau C, Jozsa R, Peres A and Wootters W K 1993 \textit{Phys. Rev. Lett.} \textbf{70}(13): 1895--1899 DOI: 10.1103/PhysRevLett.70.1895

\bibitem{lab4} Song D, He C, Cao Z and Chai G 2018 \textit{IEEE Commun. Lett.} \textbf{22}(12): 2427--2430 DOI: 10.1109/LCOMM.2018.2874025

\bibitem{lab5} Verma V 2020 \textit{IEEE Commun. Lett.} \textbf{25}(3): 936--939 DOI: 10.1109/LCOMM.2020.3036587

\bibitem{lab6} \textcolor{black}{Jung E \textit{et al.} 2008 \textit{Phys. Rev. A} \textbf{78}(1): 012312 DOI: 10.1103/PhysRevA.78.012312}

\bibitem{lab7} \textcolor{black}{Dai H Y, Chen P X and Li C Z 2004 \textit{Opt. Commun.} \textbf{231}(1--6): 281--287 DOI: 10.1016/j.optcom.2003.11.074}

\bibitem{lab8} Im D G \textit{et al.} 2021 \textit{npj Quantum Inf.} \textbf{7}: 86 DOI: 10.1038/s41534-021-00426-x

\bibitem{lab9} Harraz S, Cong S and Nieto J J 2022 \textit{IEEE Commun. Lett.} \textbf{26}(3): 528--531 DOI: 10.1109/LCOMM.2021.3138854

\bibitem{lab10} \textcolor{black}{Zhang J Y, Cong S, Wang C and Harraz S 2022 \textit{Acta Phys. Sin.} \textbf{71}(22): 220303  DOI: 10.7498/aps.71.20220760}

\bibitem{lab11} \textcolor{black}{Xiao X, Yao Y, Li Y L and Xie Y M 2020 \textit{Eur. Phys. J. Plus} \textbf{135}(1): 79 DOI: 10.1140/epjp/s13360-019-00010-5}

\bibitem{lab12} Man Z X, Xia Y J and An N B 2007 \textit{Phys. Rev. A} \textbf{75}(5): 052306 DOI: 10.1103/PhysRevA.75.052306

\bibitem{lab13} Chen X B, Zhang N, Lin S, Wen Q Y and Zhu F C 2008 \textit{Opt. Commun.} \textbf{281}(8): 2331--2335 DOI: 10.1016/j.optcom.2007.12.002

\bibitem{lab14} Li X H and Ghose S 2015 \textit{Phys. Rev. A} \textbf{91}(1): 012320 DOI: 10.1103/PhysRevA.91.012320

\bibitem{lab15} Hou K, Bao D Q, Zhu C J and Yang Y P 2019 \textit{Quantum Inf. Process.} \textbf{18}(4): 104 DOI: 10.1007/s11128-019-2218-5

\bibitem{lab16} \textcolor{black}{Barasinski A, Cernoch A and Lemr K 2019 \textit{Phys. Rev. Lett.} \textbf{122}(17): 170501 DOI: 10.1103/PhysRevLett.122.170501}

\bibitem{lab17} \textcolor{black}{Cacciapuoti A S, Caleffi M, Van Meter R and Hanzo L 2020 \textit{IEEE Trans. Commun.} \textbf{68}(6): 3808--3833 DOI: 10.1109/TCOMM.2020.2978071}

\bibitem{lab18} \textcolor{black}{Oh S, Lee S and Lee H W 2002 \textit{Phys. Rev. A} \textbf{66}(2): 022316 DOI: 10.1103/PhysRevA.66.022316}

\bibitem{lab19} \textcolor{black}{Fortes R and Rigolin G 2015 \textit{Phys. Rev. A} \textbf{92}(1): 012338 DOI: 10.1103/PhysRevA.92.012338}

\bibitem{lab20} \textcolor{black}{Fonseca A 2019 \textit{Phys. Rev. A} \textbf{100}(6): 062311 DOI: 10.1103/PhysRevA.100.062311}

\bibitem{lab21} \textcolor{black}{Bennett C H, Brassard G, Popescu S, Schumacher B, Smolin J A and Wootters W K 1996 \textit{Phys. Rev. Lett.} \textbf{76}(5): 722--725 DOI: 10.1103/PhysRevLett.76.722}

\bibitem{lab22} \textcolor{black}{Krastanov S, Albert V V and Jiang L 2019 \textit{Quantum} \textbf{3}: 123 DOI: 10.22331/q-2019-02-18-123}

\bibitem{lab23} \textcolor{black}{Li Z D \textit{et al.} 2020 \textit{Phys. Rev. Res.} \textbf{2}(2): 023047 DOI: 10.1103/PhysRevResearch.2.023047}

\bibitem{lab24} \textcolor{black}{Shen L T, Chen R X, Yang Z B, Wu H Z and Zheng S B 2014 \textit{Opt. Lett.} \textbf{39}(20): 6046--6049 DOI: 10.1364/OL.39.006046}

\bibitem{lab25} Cramer J \textit{et al.} 2016 \textit{Nat. Commun.} \textbf{7}: 11526 DOI: 10.1038/ncomms11526

\bibitem{lab26} \textcolor{black}{Ofek N \textit{et al.} 2016 \textit{Nature} \textbf{536}(7617): 441--445 DOI: 10.1038/nature18949}

\bibitem{lab27} Hammond A M, Frank I W and Camacho R M 2018 \textit{IEEE J. Sel. Topics Quantum Electron.} \textbf{24}(6): 3900308 DOI: 10.1109/JSTQE.2018.2846024

\bibitem{lab28} Beale S J, Wallman J J, Gutierrez M, Brown K R and Laflamme R 2018 \textit{Phys. Rev. Lett.} \textbf{121}(19): 190501 DOI: 10.1103/PhysRevLett.121.190501

\bibitem{lab29} Grassl M, Kong L, Wei Z, Yin Z Q and Zeng B 2018 \textit{IEEE Trans. Inf. Theory} \textbf{64}(6): 4674--4685 DOI: 10.1109/TIT.2018.2790423

\bibitem{lab30} \textcolor{black}{David F L, Lorenzo C and Markus M 2023 \textit{Quantum} \textbf{7}: 942 DOI: 10.22331/q-2023-03-09-942}

\bibitem{lab31} \textcolor{black}{Chandra D, Cacciapuoti A S, Caleffi M and Hanzo L 2022 \textit{IEEE Trans. Commun.} \textbf{70}(1): 469--484 DOI: 10.1109/TCOMM.2021.3122786}

\bibitem{lab32} \textcolor{black}{Harraz S, Cong S and Nieto J J 2022 \textit{Int. J. Quantum Inf.} \textbf{20}(4): 2250007 DOI: 10.1142/S0219749922500071}

\bibitem{lab33} \textcolor{black}{Gregoratti M and Werner R F 2003 \textit{J. Mod. Opt.} \textbf{50}(6--7): 915--933 DOI: 10.1080/0950034021000058021}

\bibitem{lab34} \textcolor{black}{Wang K, Zhao X and Yu T 2014 \textit{Phys. Rev. A} \textbf{89}(4): 042320 DOI: 10.1103/PhysRevA.89.042320}

\bibitem{lab35} \textcolor{black}{Wu H J, Jin Z and Zhu A D 2018 \textit{Int. J. Theor. Phys.} \textbf{57}(4): 1235--1244 DOI: 10.1007/s10773-017-3653-7}

\bibitem{lab36} \textcolor{black}{Xu X M, Cheng L Y, Liu A P, Su S L, Wang H F and Zhang S 2015 \textit{Quantum Inf. Process.} \textbf{14}(11): 4147--4162 DOI: 10.1007/s11128-015-1111-0}

\bibitem{lab37} \textcolor{black}{Li Y L, Sun F, Yang J and Xiao X 2021 \textit{Quantum Inf. Process.} \textbf{20}(2): 55 DOI: 10.1007/s11128-021-02998-1}

\bibitem{lab38} \textcolor{black}{Zhao X, Hedemann S R and Yu T 2013 \textit{Phys. Rev. A} \textbf{88}(2): 022321 DOI: 10.1103/PhysRevA.88.022321}

\bibitem{lab39} Agrawal P and Pati A 2006 \textit{Phys. Rev. A} \textbf{74}(6): 062320 DOI: 10.1103/PhysRevA.74.062320

\bibitem{lab40} \textcolor{black}{Nielsen M A and Chuang I L 2010 \textit{Quantum Computation and Quantum Information} New York: Cambridge University Press}

\bibitem{lab41} \textcolor{black}{Nakahara M and Ohmi T 2008 \textit{Quantum Computing: From Linear Algebra to Physical Realizations} Boca Raton: CRC Press}

\end{thebibliography}
\end{document}